\documentclass[11pt,onecolumn]{IEEEtran}
\usepackage[T1]{fontenc}
\usepackage[latin9]{inputenc}
\usepackage[cmex10]{amsmath}
\usepackage{amsthm}
\usepackage{amstext}
\usepackage{amssymb}
\usepackage{graphicx}
\usepackage[unicode=true,pdfusetitle,
 bookmarks=true,bookmarksnumbered=true,bookmarksopen=false,
 breaklinks=false,pdfborder={0 0 1},backref=false,colorlinks=false]
 {hyperref}

\makeatletter

\providecommand{\tabularnewline}{\\}
\newcommand{\lyxdot}{.}

\theoremstyle{plain}
\newtheorem{thm}{\protect\theoremname}
\theoremstyle{plain}
\newtheorem{lem}{\protect\lemmaname}
\theoremstyle{definition}
\newtheorem{defn}{\protect\definitionname}
\theoremstyle{definition}
\newtheorem{example}{\protect\examplename}
\theoremstyle{plain}

\theoremstyle{remark}
\newtheorem{rem}{\protect\remarkname}
\theoremstyle{plain}
\newtheorem{cor}{\protect\corollaryname}

\usepackage[table]{xcolor}

\newcommand{\spanset}{\mathop{\mathrm{span}}}
\newcommand{\diag}{\mathop{\mathrm{diag}}}

\newcommand{\vv}{\mathbf{v}}
\newcommand{\vu}{\mathbf{u}}
\newcommand{\vx}{\mathbf{x}}
\newcommand{\vb}{\mathbf{b}}
\newcommand{\vy}{\mathbf{y}}
\newcommand{\vs}{\mathbf{s}}
\newcommand{\vz}{\mathbf{z}}

\newcommand{\mA}{\mathbf{A}}

\@ifundefined{showcaptionsetup}{}{%
 \PassOptionsToPackage{caption=false}{subfig}}
\usepackage{subfig}
\makeatother

\providecommand{\corollaryname}{Corollary}
\providecommand{\definitionname}{Definition}
\providecommand{\examplename}{Example}
\providecommand{\factname}{Fact}
\providecommand{\lemmaname}{Lemma}
\providecommand{\remarkname}{Remark}
\providecommand{\theoremname}{Theorem}

\begin{document}

\title{Diophantine Approach to Blind Interference Alignment of
Homogeneous $K$-user $2\times1$ MISO Broadcast Channels}

\author{Qing F.~Zhou {\em Member, IEEE} , Q. T. Zhang, {\em Fellow,
IEEE} and Francis C. M. Lau, {\em Senior Member, IEEE}
\thanks{Qing F.~Zhou and Q. T. Zhnag are with the Department of Electronic Engineering,
City University of Hong Kong, Kowloon, Hong Kong.
Francis C.~M.~Lau is with the Department of Electronic
and Information Engineering, The Hong Kong Polytechnic University,
Kowloon, Hong Kong. (Email: enqfzhou@ieee.org, wirelessqt@gmail.com,
encmlau@polyu.edu.hk, ).}
\thanks{This work was supported by City University of Hong Kong under xxxxxxx. } }

\maketitle


\begin{abstract}

Although the sufficient condition for a blindly interference-aligned (BIA) 2-user $2\times 1$ broadcast channel (BC) in homogeneous fading to achieve its maximal 4/3 DoF is   well understood, its counterpart for the general $K$-user $2\times 1$ MISO BC in homogeneous block fading to achieve the corresponding $\tfrac{2K}{2+K-1}$ (DoF) remains unsolved and is, thus, the focus of this paper. An interference channel is said BIA-feasible if it achieves its maximal DoF only via BIA. In this paper, we cast this general feasibility problem in the framework of finding integer solutions for a system of linear Diophantine equations. By assuming independent user links each of the same coherence time and by studying the solvability of the Diophantine system, we derive the sufficient and necessary conditions on the $K$ users' fading block offsets to ensure the BIA feasibility of the $K$-user BC. If the $K$ offsets are independent and uniformly distributed over a coherence block, we can further prove that 11 users are enough for one to find, with certainty of $95\%$, 3 users among them to form a BIA-feasible $3$-user $2\times 1$ BC.
\end{abstract}



\section{Introduction }
Degree-of-freedom (DoF), as a more tractable performance measure than capacity region, has been widely studied to characterize lots of communication channels, such as Gaussian interference channel. Briefly speaking, DoF represents the slope of the \emph{asymptotic} achievable rate as the signal-to-noise ratio (SNR) approaches infinity. It is an equivalent measure to multiplexing gain, geometrically signifying the interference-free signal dimensions, for instance, in the context of multi-input multi-output (MIMO) channels. Interference Alignment (IA), as a powerful signal processing method in communication system \cite{Cadambe2008,Gou2011,Gou2012,Jafar2012}, was discovered recently when studying the maximal achievable DoF for X channels \cite{Maddah-Ali2006,Jafar2006}, and for multi-input single-output (MISO) compound broadcast channels (BC) \cite{Weingarten2007}. The most surprising result by IA is that, by carefully signaling design on the transmitters, the total DoF achievable at a $K$-user interference channel reaches $K/2$, considerably higher than the previous belief of the maximal 1 DoF, which is achievable by orthogonal interference scheduling.

Based on the extent of the channel state information known at the transmitters (CSIT), the implementation of IA is able to be categorized in three types, namely \emph{IA with perfect CSIT}, \emph{IA with delayed CSIT} and \emph{IA with no need of CSIT}. The method of IA with perfect CSIT can be further divided into the signal vector space method \cite{Cadambe2008,Maddah-Ali2008,Yetis2010}, and the signal scaling method \cite{Etkin2009,Motahari2009}. However, in practical systems, instantaneous and perfect CSIT is beyond reality. Technically speaking, the DoF region obtained by IA with perfect CSIT only serves as an upper bound on achievable DoF.

Fortunately, when imperfect CSIT is available, IA is still able to be implemented and provide DoF gain. It is shown \cite{Maddah-Ali2010, Maddah-Ali2012} that, in the context of a compound 2-user $2\times 1$ MISO BC, the outer-bounded 4/3 DoF is achievable by IA, even possessing only delayed (outdated/stale) CSIT. By contrast, DoF 1 is optimal for the MISO BC if IA is not applied. This method of IA with only need of delayed CSIT is also proved suitable for distributed transmitters \cite{Maleki2012,Abdoli2011,Vaze2012}.


Surprisingly, for the 2-user $2\times 1$ MISO BC mentioned above, the 4/3 DoF can still be achieved by IA, even with no knowledge of CSIT \cite{Gou2011,Jafar2012}. But the BC needs to meet two requirements, one is that it has only a finite number of time slots, and the other is that the finite-time-slot BC has certain staggered channel matrix structure \cite{Gou2011,Jafar2012}. The special channel matrix structure over finite time slots is either generated artificially \cite{Gou2011} or found in certain heterogeneous block fading cases \cite{Jafar2012}. This IA method with no need of CSIT is usually referred to as blind interference alignment (BIA). The essential idea of BIA is that a symbol $x$ is transmitted twice, due to the structured channel matrix, at the desired user it is received as $h_1x$ and $g_1x$, respectively, while at an undesired user the received signals are $h_ix$ and $h_ix$. Here, $h$ and $g$ denote channel coefficients. By subtracting the second received signal from the first received one, the interfering signal $x$ is removed ($h_ix-h_ix = 0$) from the undesired user, while $(h_1-g_1)x$ is left for further process at the desired user. It is recently shown that BIA also provides significant DoF gain for cellular networks  \cite{Wang2011,Wang2011a,Jafar2012a}, which can be viewed as an interference network with partial connectivity. More information-theoretical study on the DoF gain concerning the BIA method can be found in \cite{Huang2012,Zhu2012, Gou2011a} and the references therein.

Unlike the IA method with the need of perfect CSIT or delayed CSIT, BIA needs no overhead for feedback to gain CSI at the transmitters, incurring no delay and complexity, and thus is easy to be incorporated in existing communication systems and of practical interest in the advance of modern communication. Concerning BIA, prior works  \cite{Gou2011,Jafar2012} mainly focused on, for instance in the context of $K$-user $L\times 1$ MISO BC, finite channel block and heterogeneous block fading. It is still not well studied whether the optimal DoF $\tfrac{LK}{L+K-1}$ can be achievable, by using BIA, in a more general $K$-user $L\times 1$ MISO BC setting, such as homogeneous block fading over infinite time slots.
By homogeneous block fading, we mean that the links connecting the transmitter and the users, undertaking independent block fading, have an identical coherence time.

Our recent preliminary work \cite{Zhou2012} shows that, in a  homogeneous 2-user $2\times 1$ MISO BC, by default spanning infinite time slots in this paper, the optimal 4/3 DoF is achievable by using BIA, as long as the relative offset of the two users' fading blocks falls in the range $[\lceil \tfrac{N}{3}\rceil, \lfloor \tfrac{2N}{3}\rfloor]$, where $N$ is the coherence time. This result contains the finding of \cite{Gou2012,Jafar2012}, in which $N=2$ and two users' fading blocks are staggered, as a special case. To show the achievability in that paper, we first identify all channel patterns of BIA-feasible super-symbol channel block, which contains three time slots and is able to convey four symbols by using BIA, resulting in 4/3 DoF. We then prove a homogenous 2-user $2\times 1$ BC, if meeting the sufficient condition above, can be completely decomposed into small BIA-feasible channel blocks. The method is difficult, however, to be extended to homogeneous $K$-user $2\times 1$ MISO BC, since the presentative matrix used in the method has the size of $K\times KN$, which incurs polynomially increasing complexity on $K$ and $N$. Moreover, the method cannot prove the derived sufficient condition is the necessary one.

In this paper, we apply new methodology to address the BIA-feasibility problem for a general $K$-user $2\times 1$ MISO BC with homogeneous fading. We say such a MISO BC is BIA-feasible if, by using BIA, the optimal $\tfrac{2K}{2+K-1}$ DoF \cite{Gou2012,Jafar2012} can be achieved over the infinite-time-slot channel. To completely characterize the MISO BC, the coherence time $N$ and the fading block offset $(n_{\delta,1},\cdots,n_{\delta,K})$ are applied. Rather than use the prior complicated method in \cite{Zhou2012}, we first exploit a simple method, which applies a $12\times 12$ representative matrix, for $K=3$. Then, for the BCs with $K>3$, we extend the method and further cast the BIA-feasibility problem into the solvability problem of a system of linear equations with all variables being integers, mathematically known as a system of linear Diophantine equations \cite{Lazebnik1996}. Unlike the general linear Diophantine system $\mA\vx=\vb$, whose solvability is normally difficulty to determine by only examining the algebraic structure of $\vb$ \cite{Lazebnik1996, Chou1982}, we will show that the solvability problem of the linear system in this paper can be completely determined by looking into the structure of $(n_{\delta,1},\cdots,n_{\delta,K})$.
This paper presents four main contributions. Firstly, for the 3-user BC we prove the sufficient and necessary BIA-feasible condition on $n_{\delta,k}$s. Secondly, we derive the probability of finding three users to form a BIA-feasible 3-user BC from a group of $K\geq 3$ users, whose offsets are independently and uniformly placed over $[0,N-1]$. Thirdly, we generalize the result for the 3-user BC and derive the sufficient and necessary BIA-feasible condition for the $K$-user BC. Finally, it can be concluded that, when $K$ goes large, there exists $k<K$ such that a BIA-feasible $k$-user $2\times 1$ BC can be found for sure, asymptotically achieving the maximal DoF 2 and forming a virtual $2\times k$ MIMO channel.

{\bf Notations:} Throughout this paper, vectors are represented by lower case bold font, like $\vu$, $\vv$; matrices are represented by upper case bold font, like $\mA$. Exception is made on the representations
for channel coefficients, in particular, $H_i$ represents channel coefficient vector while $H_{ij}$ represents channel coefficient matrix.

\section{System model}
\noindent \begin{center}
\begin{figure}
\noindent \begin{centering}
\includegraphics[clip,scale=0.6]{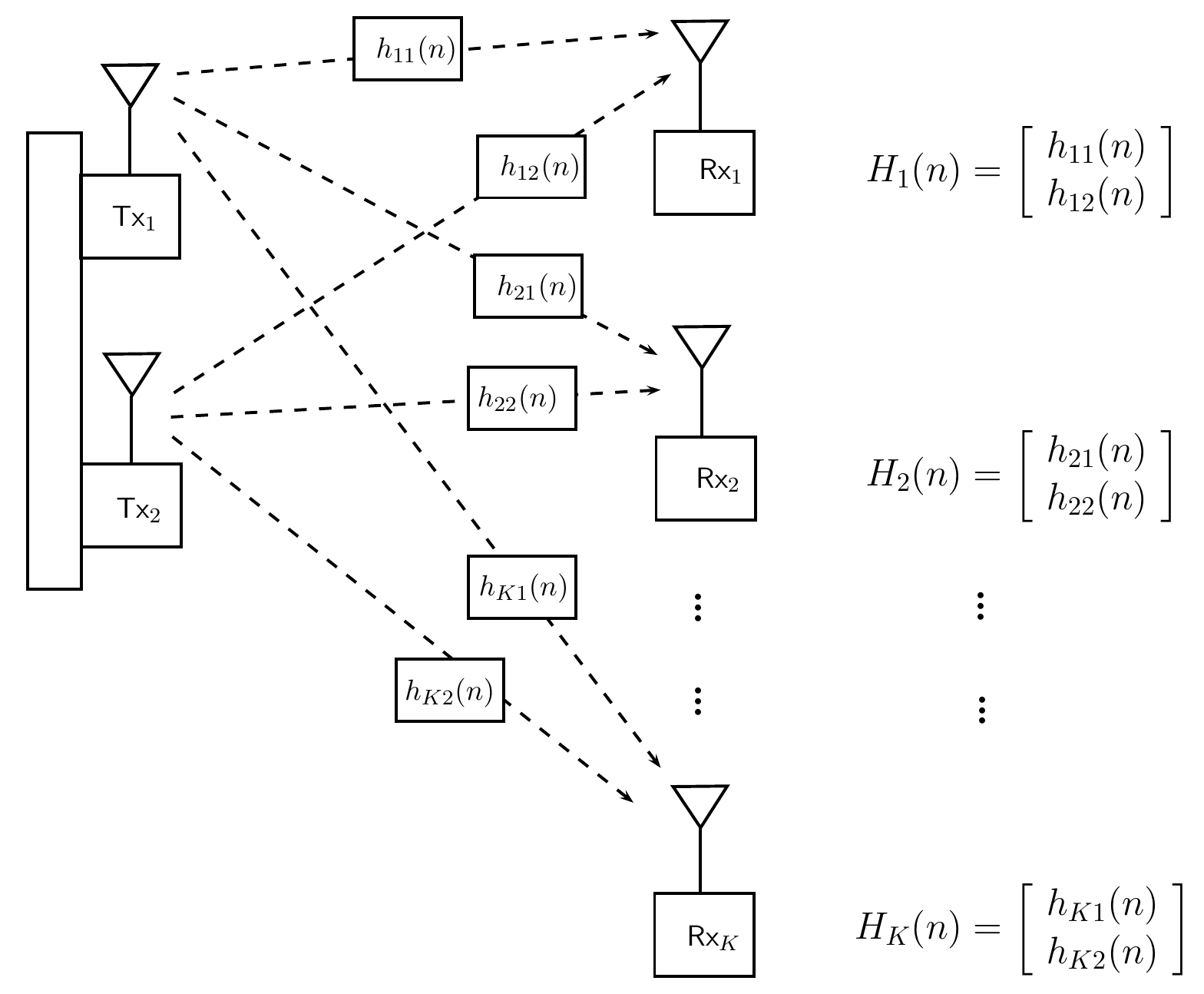}
\par\end{centering}
\caption{System model of $K$-user $2\times1$ MISO broadcast channel.}
\label{fig:system.model}
\end{figure}
\par\end{center}

Consider a $K$-user $2\times1$ MISO broadcast channel (BC) in which, a
transmitter with 2 antennas broadcasts $K$ independent signals to $K$ users, through a respective $2\times1$ MISO array of antennas. As sketched in
Fig. \ref{fig:system.model}, the transmitter employs two antennas, $\mathsf{Tx}_{1}$
and $\mathsf{Tx}_{2}$, to convey $K$ parallel information streams,
respectively, to the $K$ single-antenna users, denoted as $\mathsf{Rx}_{i}$,
$i\in\{1,\cdots,K\}$. Let $h_{ij}(n)$ denote the channel response, in baseband form, for the link from $\mathsf{Tx}_{j}$ to $\mathsf{Rx}_{i}$ at the time $n$. We can write the channel coefficients linking user $i$ to the transmit antennas, in vector form, as $H_{i}(n)=[h_{i1}(n),h_{i2}(n)]^{T}$. The BC considered in this paper spans infinite time slots, that is, $n\in\{0,\cdots,\infty\}$, unless explicitly stated otherwise.

Furthermore, we consider homogeneous $K$-user $2\times 1$ BC, in which the $K$ $2\times1$ MISO links experience independent block fading with an identical coherence time $N$. Denote $N_{i}$ as the coherence time for $H_{i}(n)$ seen at the user
$\mathsf{Rx}_{i}$, and $n_{\delta,i}$ as the initial time offset.
Then we have $N_i = N$ for all $i\in\{1,\cdots,K\}$. In addition, we further assume $0\leq n_{\delta,i}<N$.

For the aforementioned broadcasting channel without assuming channel
state information available at the transmitter side (no CSIT), it is shown \cite{Gou2011} that the optimal sum DoF is $\frac{2K}{2+K-1}$  or equivalently, $\frac{2}{K+1}$
DoF for each user. As comparison,  the maximal sum DoF with perfect CSIT is 2, which is
achievable by using beamforming. In the following, we will analyze the sufficient and necessary conditions on the time offsets $n_{\delta, i}$ such that a general homogeneous $K$-user $2\times 1$ MISO BC, which is supposed to span infinite time slots, can achieve the optimal $\frac{2K}{2+K-1}$ DoF by using IA with no need of CSIT, that is, BIA. Such a homogeneous BC is said to be \emph{BIA-feasible}.

\section{Preliminary results for homogeneous 3-user $2\times 1$ MISO BC}

We start with the simplest case $K=3$. First we review the super-symbol channel block in \cite{Gou2011}, whose channal matrix has certain structure pattern such that the optimal $\tfrac{3}{2}$ is achievable by using BIA. Such a channel block is referred to as BIA-feasible channel block, and the structured channel pattern is referred to as BIA-feasible channel pattern. We then present BIA-feasible channel patterns which is able to be found in homogeneous $3$-user $2\times 1$ BC.

\begin{figure}
\begin{centering}
\includegraphics[scale=0.7]{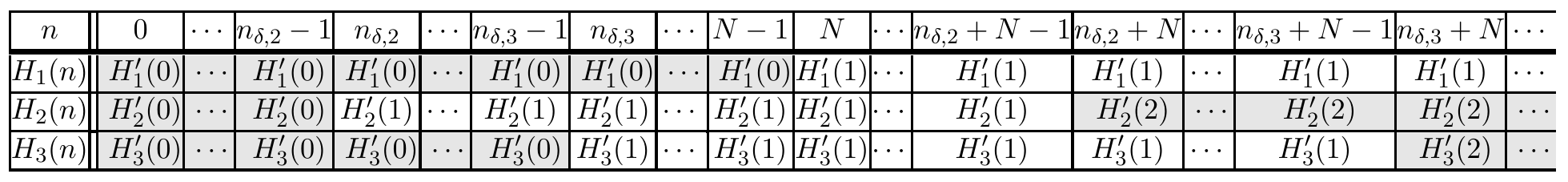}
\par\end{centering}
\caption{Channel matrix for a homogeneous 3-user $2\times1$ MISO BC channel with coherence time $N$ and $n_{\delta,1}=0$. }
\label{fig:chanel.matrix.K3}
\end{figure}

Without loss of generality, we assume $n_{\delta,1}=0$.
Fig. \ref{fig:chanel.matrix.K3} illustrates the channel matrix for the homogeneous $3$-user BC. The user $\mathsf{Rx}_{1}$ observes
the channel state $H_{1}(a_{1}N+b_{1})=H_{1}'(a_{1})$ for all $0\leq a_{1}$
and $0\leq b_{1}<N$. The other two users $\mathsf{Rx}_{i}$ ($i\in\{2,3\}$)
observe the channel state $H_{i}(b_{i})=H_{i}'(0)$ for $0\leq b_{i}\leq(n_{\delta,i}-1)$,
and $H_{i}(a_{i}N+n_{\delta,i}+b_{i})=H_{i}'(a_{i}+1)$ for all $0\leq a_{i}$
and $0\leq b_{i}<N$.

\subsection{Review of Blind Interference Alignment (BIA)}
\begin{figure}
\begin{centering}
\includegraphics[scale=0.8]{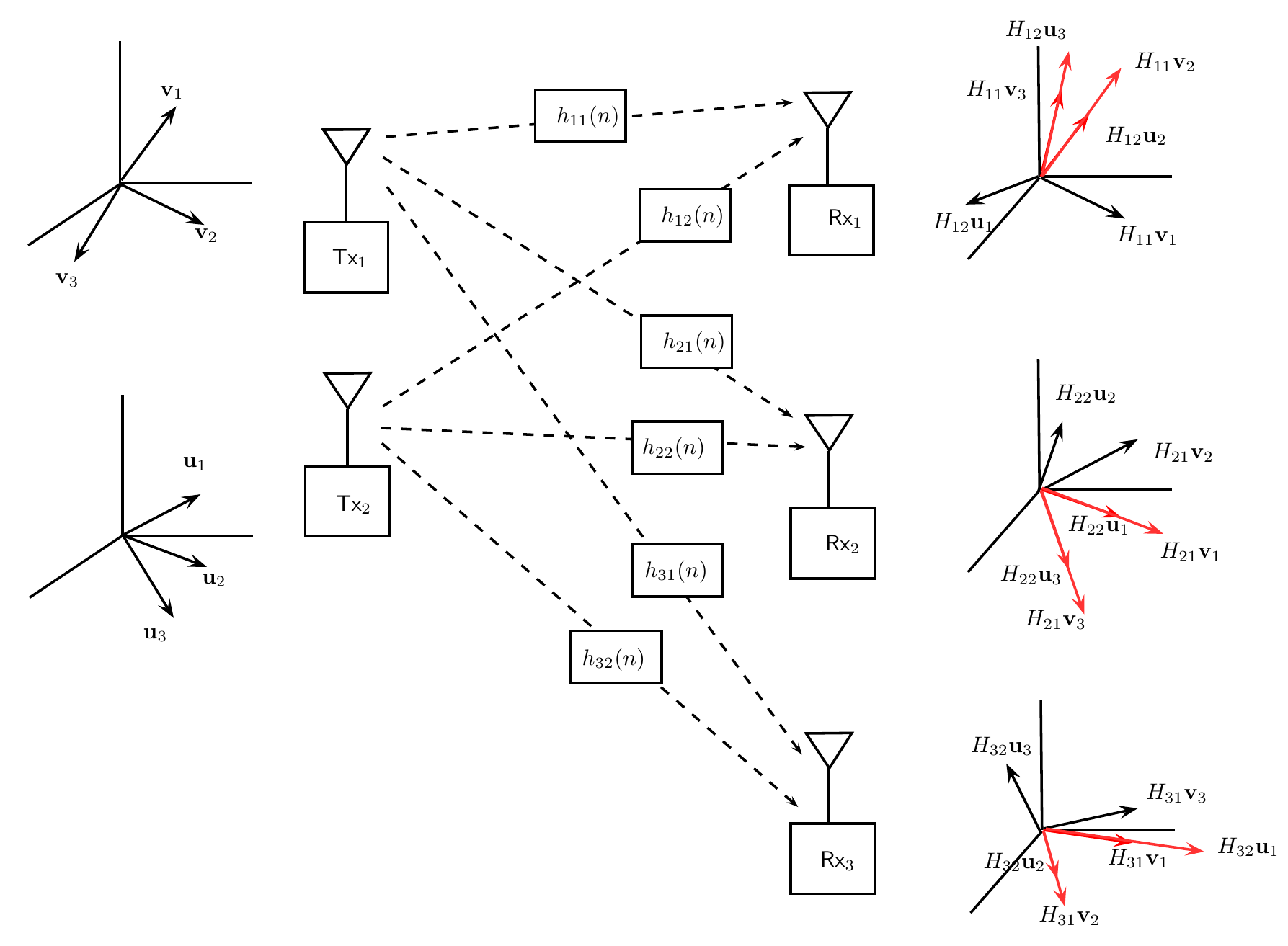}
\par\end{centering}
\caption{Interference align of the 3-user $2\times1$ MISO BC channel. Note that the symbol streams $s_{ij}$ are omitted for simplicity.}
\label{fig:IA.3u.2x1}
\end{figure}
Fig. \ref{fig:IA.3u.2x1} shows the mechanism of
interference alignment to achieves the maximal DoF $\frac{3}{2}$. We consider four time slots $n_{1}$, $n_{2}$, $n_{3}$ and $n_{4}$, which
are not necessary to be consecutive time slots. At the transmit antenna
$\mathsf{Tx}_{1}$, three signalling vectors which convey information
data are represented as $\vv_{i}=[v_{i}(n_{1}),v_{i}(n_{2}),v_{i}(n_{3}),v_{i}(n_{4})]^{T}\in\mathcal{C}^{4}$,
$i=1,2,3$. At $\mathsf{Tx}_{2}$, the three signalling vectors are
$\vu_{i}=[u_{i}(n_{1}),u_{i}(n_{2}),u_{i}(n_{3}),u_{i}(n_{4})]^{T}\in\mathcal{C}^{4}$,
$i=1,2,3$. The channel state information of the link from $\mathsf{Tx}_{j}$
to $\mathsf{Rx}_{i}$ spanning the four symbols is denoted as $H_{ij}=\diag[h_{ij}(n_{1}),h_{ij}(n_{2}),h_{ij}(n_{3}),h_{ij}(n_{4})]$, where $h_{ij}\in\mathcal{C}$ and its magnitude is lower bounded by nonzero value. At the receiver
$\mathsf{Rx}_{i}$, the received signal $\vy_i=[y_i(n_1),y_i(n_2),y_i(n_3),y_i(n_4)]^T$ is
\begin{equation}
\vy_i = H_{1i}[\vv_1,\vv_2,\vv_3]\vs_1 + H_{2i}[\vu_1,\vu_2,\vu_3]\vs_2+\vz_i
\end{equation}
where $\vs_j=[s_{j1},s_{j2},s_{j3}]^T\in \mathcal{C}^{3\times 1}$ represents three symbol streams from $\mathsf{Tx}_{j}$, and $\vz_i\in \mathcal{C}^{3\times 1}$ is the AWGN vector at $\mathsf{Rx}_i$. To achieve the optimal DoF $\frac{3}{2}$, the IA implementation shown in
Fig.~\ref{fig:IA.3u.2x1} has the following requirements.
At the user $\mathsf{Rx}_{1}$, $H_{12}\vu_{3}$ is aligned with $H_{11}\vv_{3}$,
and $H_{12}\vu_{2}$ is aligned with $H_{11}\vv_{2}$, such that $\vu_{1}$
and $\vv_{1}$ can be detected with no interference. The alignment
condition can be mathematically represented as
\begin{equation}
\begin{cases}
\spanset\{H_{12}\vu_{3}\} & =\spanset\{H_{11}\vv_{3}\}\\
\spanset\{H_{12}\vu_{2}\} & =\spanset\{H_{11}\vv_{2}\}
\end{cases}.\label{eq:cond1}
\end{equation}
Similarly, the user $\mathsf{Rx}_{2}$ detects $\vv_{2}$ and $\vu_{2}$
with no interference by demanding the alignment condition
\begin{equation}
\begin{cases}
\spanset\{H_{21}\vv_{1}\} & =\spanset\{H_{22}\vu_{1}\}\\
\spanset\{H_{21}\vv_{3}\} & =\spanset\{H_{22}\vu_{3}\}
\end{cases}.\label{eq:cond2}
\end{equation}
The user $\mathsf{Rx}_{3}$ detects $\vv_{3}$ and $\vu_{3}$ and
demanding the alignment condition
\begin{equation}
\begin{cases}
\spanset\{H_{31}\vv_{1}\} & =\spanset\{H_{32}\vu_{1}\}\\
\spanset\{H_{31}\vv_{2}\} & =\spanset\{H_{32}\vu_{2}\}
\end{cases}\label{eq:cond3}
\end{equation}
In this implementation, $\mathsf{Rx}_i$ decodes the symbols delivered by $\vv_i$ and $\vu_i$, i.e., the symbols $s_{1i}$ from $\mathsf{Tx}_1$ and $s_{2i}$ from $\mathsf{Tx}_2$. Six symbols are delivered after four channel uses, so the DoF $\tfrac{3}{2}$ is achieved.
The alignment conditions \eqref{eq:cond1}, \eqref{eq:cond2}
and \eqref{eq:cond3} can be rewritten as
\begin{equation}
\begin{cases}
\vv_{1} & \rightarrow\diag\left[\frac{h_{22}(n_{1})}{h_{21}(n_{1})},\frac{h_{22}(n_{2})}{h_{21}(n_{2})},\frac{h_{22}(n_{3})}{h_{21}(n_{3})},\frac{h_{22}(n_{4})}{h_{21}(n_{4})}\right]\vu_{1}\rightarrow\diag\left[\frac{h_{32}(n_{1})}{h_{31}(n_{1})},\frac{h_{32}(n_{2})}{h_{31}(n_{2})},\frac{h_{32}(n_{3})}{h_{31}(n_{3})},\frac{h_{32}(n_{4})}{h_{31}(n_{4})}\right]\vu_{1}\\
\vv_{2} & \rightarrow\diag\left[\frac{h_{12}(n_{1})}{h_{11}(n_{1})},\frac{h_{12}(n_{2})}{h_{11}(n_{2})},\frac{h_{12}(n_{3})}{h_{11}(n_{3})},\frac{h_{12}(n_{4})}{h_{11}(n_{4})}\right]\vu_{2}\rightarrow\diag\left[\frac{h_{32}(n_{1})}{h_{31}(n_{1})},\frac{h_{32}(n_{2})}{h_{31}(n_{2})},\frac{h_{32}(n_{3})}{h_{31}(n_{3})},\frac{h_{32}(n_{4})}{h_{31}(n_{4})}\right]\vu_{2}\\
\vv_{3} & \rightarrow\diag\left[\frac{h_{12}(n_{1})}{h_{11}(n_{1})},\frac{h_{12}(n_{2})}{h_{11}(n_{2})},\frac{h_{12}(n_{3})}{h_{11}(n_{3})},\frac{h_{12}(n_{4})}{h_{11}(n_{4})}\right]\vu_{3}\rightarrow\diag\left[\frac{h_{22}(n_{1})}{h_{21}(n_{1})},\frac{h_{22}(n_{2})}{h_{21}(n_{2})},\frac{h_{22}(n_{3})}{h_{21}(n_{3})},\frac{h_{22}(n_{4})}{h_{21}(n_{4})}\right]\vu_{3}
\end{cases},\label{eq:sig.vec.cond}
\end{equation}
where $\vv\rightarrow\vu$ means that $\vv=a\vu$ for a non-zero scale
$a$.

Over the four time slots, when the channel matrix $H_{ij}(n)$ fits certain structures, $\vv_i$ and $\vu_i$ are able to be chosen independent of the value of $H_{ij}(n)$. This kind of interference alignment implementation with no need of CSIT is called BIA. Note that the alignment conditions demonstrated by this example is
stricter than the general IA conditions, which
are simply $\spanset\{H_{i1}\vv_{j}:j\neq i,j\in\{1,2,3\}\}=\spanset\{H_{i2}\vu_{j}:j\neq i,j\in\{1,2,3\}\}$
for $i=1,2,3$. Along our analysis, we will show that when BIA is concerned the general IA conditions is about to degenerate to the form of the conditions \eqref{eq:cond1}, \eqref{eq:cond2}
and \eqref{eq:cond3}.

\subsection{BIA-feasible super-symbol channel blocks}

According to priori works \cite{Gou2011,Jafar2012}, a super-symbol,
which is composed of four time slots $n_{i}$, $i\in\{1,2,3,4\}$, is BIA-feasible
if it has the following channel state pattern

\begin{center}
\begin{tabular}{|c|c|c|c|}
\hline
$H_{1}'(\beta)$ & \cellcolor[gray]{0.9}$H_{1}'(\alpha)$ & $H_{1}'(\beta)$ & $H_{1}'(\beta)$\tabularnewline
\hline
$H_{2}'(\gamma)$ & $H_{2}'(\gamma)$ & \cellcolor[gray]{0.9}$H_{2}'(\psi)$ & $H_{2}'(\gamma)$\tabularnewline
\hline
$H_{3}'(\rho)$ & $H_{3}'(\rho)$ & $H_{3}'(\rho)$ & \cellcolor[gray]{0.9}$H_{3}'(\pi)$\tabularnewline
\hline
\end{tabular}.
\par\end{center}
The BIA-feasible super-symbol channel pattern
above, however, would not appear in any homogeneous BC since the channel
state sequence at the user $\mathsf{Rx}_{1}$ ----- $H_{1}'(\beta)$,
$H_{1}'(\alpha)$, $H_{1}'(\beta)$, $H_{1}'(\beta)$ ----- doesn't
meet the block fading premise. In homogeneous BC, the following 4-symbol
channel pattern is BIA-feasible

\begin{center}
\begin{tabular}{|c|c|c|c|}
\hline
\cellcolor[gray]{0.9}$H_{1}'(\alpha)$ & \cellcolor[gray]{0.9}$H_{1}'(\alpha)$ & $H_{1}'(\beta)$ & $H_{1}'(\beta)$\tabularnewline
\hline
\cellcolor[gray]{0.9}$H_{2}'(\gamma)$ & $H_{2}'(\psi)$ & $H_{2}'(\psi)$ & $H_{2}'(\psi)$\tabularnewline
\hline
\cellcolor[gray]{0.9}$H_{3}'(\rho)$ & \cellcolor[gray]{0.9}$H_{3}'(\rho)$ & \cellcolor[gray]{0.9}$H_{3}'(\rho)$ & $H_{3}'(\pi)$\tabularnewline
\hline
\end{tabular}
\par\end{center}

\noindent \begin{flushleft}
To see its feasibility, we substitute the channel state into \eqref{eq:sig.vec.cond}
and get the explicit alignment conditions
\begin{equation}
\begin{cases}
\vv_{1} & \rightarrow\diag\left[\frac{h_{22}'(\gamma)}{h_{21}'(\gamma)},\frac{h_{22}'(\psi)}{h_{21}'(\psi)},\frac{h_{22}'(\psi)}{h_{21}'(\psi)},\frac{h_{22}'(\psi)}{h_{21}'(\psi)}\right]\vu_{1}\rightarrow\diag\left[\frac{h_{32}'(\rho)}{h_{31}'(\rho)},\frac{h_{32}'(\rho)}{h_{31}'(\rho)},\frac{h_{32}'(\rho)}{h_{31}'(\rho)},\frac{h_{32}'(\pi)}{h_{31}'(\pi)}\right]\vu_{1}\\
\vv_{2} & \rightarrow\diag\left[\frac{h_{12}'(\alpha)}{h_{11}'(\alpha)},\frac{h_{12}'(\alpha)}{h_{11}'(\alpha)},\frac{h_{12}'(\beta)}{h_{11}'(\beta)},\frac{h_{12}'(\beta)}{h_{11}'(\beta)}\right]\vu_{2}\rightarrow\diag\left[\frac{h_{32}'(\rho)}{h_{31}'(\rho)},\frac{h_{32}'(\rho)}{h_{31}'(\rho)},\frac{h_{32}'(\rho)}{h_{31}'(\rho)},\frac{h_{32}'(\pi)}{h_{31}'(\pi)}\right]\vu_{2}\\
\vv_{3} & \rightarrow\diag\left[\frac{h_{12}'(\alpha)}{h_{11}'(\alpha)},\frac{h_{12}'(\alpha)}{h_{11}'(\alpha)},\frac{h_{12}'(\beta)}{h_{11}'(\beta)},\frac{h_{12}'(\beta)}{h_{11}'(\beta)}\right]\vu_{3}\rightarrow\diag\left[\frac{h_{22}'(\gamma)}{h_{21}'(\gamma)},\frac{h_{22}'(\psi)}{h_{21}'(\psi)},\frac{h_{22}'(\psi)}{h_{21}'(\psi)},\frac{h_{22}'(\psi)}{h_{21}'(\psi)}\right]\vu_{3}
\end{cases}.\label{eq:ex.pattern}
\end{equation}
It is easy to see that the conditions are satisfied if we choose $\vv_{1}=\vu_{1}=[0,1,1,0]^{T}$,
$\vv_{2}=\vu_{2}=[1,1,0,0]^{T}$ and $\vv_{3}=\vu_{3}=[0,0,1,1]^{T}$,
and thus BIA is achieved.
\par\end{flushleft}

We then study the number of BIA-feasible super-symbol channel patterns
with 4-time symbol extension, in the homogeneous BC, by proving the following lemma.
\begin{lem}
\label{lem:patternMatrix}In homogeneous $3$-user $2\times1$ MISO BCs, there
are $3!=6$ super-symbol channel patterns with $4$ symbol extension
which are BIA-feasible. \end{lem}
\begin{IEEEproof}
First we prove that the conditions in the form of \eqref{eq:sig.vec.cond}
is necessary to achieve BIA. We prove this by contradiction. As shown
in Fig. \eqref{fig:IA.3u.2x1}, suppose at $\mathsf{Rx}_{1}$ $\vu_{3}$
falls in the subspace spanned by $H_{11}\vv_{2}$ and $H_{11}\vv_{3}$,
but not align with either of them. We can express this as
\begin{equation}
\vu_{3}=a\frac{H_{11}}{H_{12}}\vv_{2}+b\frac{H_{11}}{H_{12}}\vv_{3},\label{eq:span.assumption}
\end{equation}
where $a\neq0$ and $b\neq0$. Since $\vu_{3}$, $\vv_{2}$ and $\vv_{3}$
are not channel dependent, we can rewrite the equation above as
\begin{equation}
\vu_{3}\rightarrow a\vv_{2}+b\vv_{3}.
\end{equation}
Similarly, at $\mathsf{Rx}_{2}$ it must be satisfied that
\begin{equation}
\vu_{3}\rightarrow c\vv_{1}+d\vv_{3}.
\end{equation}
In this equation, it must be $d\neq0$, otherwise $\{\vv_{1},\vv_{2},\vv_{3}\}$
forms a dependent set, which contradicts the signaling vector
design at $\mathsf{Tx}_{2}$. It must be $c\neq0$, otherwise $\vu_{3}$
is aligned with $\vv_{3}$, which contradicts the assumption that
$a\neq0$ and $b\neq0$ at \eqref{eq:span.assumption}. So, we conclude
$c\neq0$ and $d\neq0$, however this also leads to the result that
$\{\vv_{1},\vv_{2},\vv_{3}\}$ is a dependent set, which is impossible.
Therefore, by the contradictions above, we finish proving that \eqref{eq:sig.vec.cond}
is a necessary condition form.

We can construct a $3\times3$ pattern matrix $A$ by defining
\begin{equation}
a_{ij}=\begin{cases}
1, & \text{if}\quad H_{i}(n_{j})\neq H_{i}(n_{j+1})\\
0, & \text{if}\quad H_{i}(n_{j})=H_{i}(n_{j+1})
\end{cases}\quad\text{for}\quad i,j\in\{1,2,3\}.
\end{equation}
From this definition, the pattern matrix of the example above, which
is characterized by \eqref{eq:ex.pattern}, is given by
\begin{equation}
A=\left[\begin{array}{ccc}
0 & 1 & 0\\
1 & 0 & 0\\
0 & 0 & 1
\end{array}\right].
\end{equation}
It is easy to prove, by using the necessary condition \eqref{eq:sig.vec.cond},
that for any 4-symbol channel pattern in homogeneous BCs, it is BIA-feasible if only
if each row of its pattern matrix has a unique $1$ element and each
column of the matrix has a unique $1$ element. Therefore, a 4-symbol channel pattern
is BIA feasible if only if its pattern matrix is a $3\times3$ permutation
matrix. There are $3!=6$ $3\times3$ permutation matrices, and thus
there are $3!$ 4-symbol BIA-feasible channel patterns, which proves
the lemma.
\end{IEEEproof}
To show how the coherence time and the offsets will
affect BIA in homogeneous $3$-user $2\times1$ BC, we give a
simple example. Consider a BC with $N=4$, $n_{\delta,2}=1$ and $n_{\delta,3}=2$;
its channel coefficients over time are shown in Fig. \ref{fig:fading.ch.N4},
in which constant channel coefficients are represented by the same
symbol. The channel fragment from $n=3$ to $n=18$, which contains
$4N=16$ consecutive time slots, can by decomposed into four BIA-feasible
4-symbol channel patterns as shown in Fig. \ref{fig:BIA.implement.N4}.
Every channel fragment containing $4N=16$ consecutive symbols afterwards
has the same pattern as the one from $n=3$ to $n=18$, and thus it
can be decomposed in the same way. Therefore, this broadcast
channel spanning infinite time slots is BIA-feasible.

\begin{figure}
\begin{centering}
\begin{tabular}{|c|c|c|c|c|c|c|c|c|c|c|c|c|c|c|c|c|c|c|c|}
\hline
$n$ & $0$ & $1$ & $2$ & $3$ & $4$ & $5$ & $6$ & $7$ & $8$ & $9$ & $10$ & $11$ & $12$ & $13$ & $14$ & $15$ & $16$ & $17$ & $18$\tabularnewline
\hline
$H_{1}(n)$ & $\Box$ & $\Box$ & $\Box$ & $\Box$ & $\bullet$ & $\bullet$ & $\bullet$ & $\bullet$ & $\triangle$ & $\triangle$ & $\triangle$ & $\triangle$ & $\bigstar$ & $\bigstar$ & $\bigstar$ & $\bigstar$ & $\circ$ & $\circ$ & $\circ$\tabularnewline
\hline
$H_{2}(n)$ & $\spadesuit$ & $\nabla$ & $\nabla$ & $\nabla$ & $\nabla$ & $\blacklozenge$ & $\blacklozenge$ & $\blacklozenge$ & $\blacklozenge$ & $\ominus$ & $\ominus$ & $\ominus$ & $\ominus$ & $\blacktriangleright$ & $\blacktriangleright$ & $\blacktriangleright$ & $\blacktriangleright$ & $\otimes$ & $\otimes$\tabularnewline
\hline
$H_{3}(n)$ & $\clubsuit$ & $\clubsuit$ & $\heartsuit$ & $\heartsuit$ & $\heartsuit$ & $\heartsuit$ & $\blacktriangleleft$ & $\blacktriangleleft$ & $\blacktriangleleft$ & $\blacktriangleleft$ & $\diamondsuit$ & $\diamondsuit$ & $\diamondsuit$ & $\diamondsuit$ & $\blacktriangledown$ & $\blacktriangledown$ & $\blacktriangledown$ & $\blacktriangledown$ & $\boxdot$\tabularnewline
\hline
\end{tabular}
\par\end{centering}
\caption{\label{fig:fading.ch.N4}A homogeneous block fading channel with $N=4$,
$n_{\delta,2}=1$ and $n_{\delta,3}=2$.}
\end{figure}

\begin{figure}
\noindent \begin{centering}
\begin{tabular}{|c|c|c|c|}
\hline
$3$ & $4$ & $5$ & $6$\tabularnewline
\hline
$\Box$ & $\bullet$ & $\bullet$ & $\bullet$\tabularnewline
\hline
$\nabla$ & $\nabla$ & $\blacklozenge$ & $\blacklozenge$\tabularnewline
\hline
$\heartsuit$ & $\heartsuit$ & $\heartsuit$ & $\blacktriangleleft$\tabularnewline
\hline
\end{tabular}\quad %
\begin{tabular}{|c|c|c|c|}
\hline
$7$ & $8$ & $9$ & $10$\tabularnewline
\hline
$\bullet$ & $\triangle$ & $\triangle$ & $\triangle$\tabularnewline
\hline
$\blacklozenge$ & $\blacklozenge$ & $\ominus$ & $\ominus$\tabularnewline
\hline
$\blacktriangleleft$ & $\blacktriangleleft$ & $\blacktriangleleft$ & $\diamondsuit$\tabularnewline
\hline
\end{tabular} \quad%
\begin{tabular}{|c|c|c|c|}
\hline
$11$ & $12$ & $13$ & $14$\tabularnewline
\hline
$\triangle$ & $\bigstar$ & $\bigstar$ & $\bigstar$\tabularnewline
\hline
$\ominus$ & $\ominus$ & $\blacktriangleright$ & $\blacktriangleright$\tabularnewline
\hline
$\diamondsuit$ & $\diamondsuit$ & $\diamondsuit$ & $\blacktriangledown$\tabularnewline
\hline
\end{tabular} \quad%
\begin{tabular}{|c|c|c|c|}
\hline
$15$ & $16$ & $17$ & $18$\tabularnewline
\hline
$\bigstar$ & $\circ$ & $\circ$ & $\circ$\tabularnewline
\hline
$\blacktriangleright$ & $\blacktriangleright$ & $\otimes$ & $\otimes$\tabularnewline
\hline
$\blacktriangledown$ & $\blacktriangledown$ & $\blacktriangledown$ & $\boxdot$\tabularnewline
\hline
\end{tabular}
\par\end{centering}
\caption{\label{fig:BIA.implement.N4}The implementation of BIA for the homogeneous
block fading channel shown in Fig. \ref{fig:fading.ch.N4}.}
\end{figure}

\section{BIA-feasibility of homogeneous $3$-user $2\times1$ MISO BC}

As demonstrated by the example mentioned previously, whether a $3$-user $2\times1$ BC is BIA-feasible is determined by the coherence time $N$, and the offsets $n_{\delta,2}$ and $n_{\delta,3}$. In this section, we study how these parameters affect the BIA feasibility of the homogeneous BC. In this section, we investigate the conditions on the coherence $N$ and offsets $n_{\delta,i}$ such that a $3$-user $2\times1$ homogeneous MISO BC is BIA-feasible.

\subsection{Pattern array}

We start this section with a lemma.
\begin{lem}
\label{lem:no.delta.align}If any pair among $n_{\delta,1}$, $n_{\delta,2}$
and $n_{\delta,3}$ are equal, then the $3$-user $2\times1$ homogeneous
BC is not BIA-feasible. \end{lem}
\begin{IEEEproof}
Without loss of generality, we suppose $n_{\delta,1}=n_{\delta,2}$,
and suppose the channel can be decomposed into BIA-feasible patterns.
First, we randomly choose a time index $n_{1}$. Then the next time
index $n_{2}$ cannot be chosen at the fading block of $H_{1}(n)$
which doesn't contain $H_{1}(n_{1})$. It is because that if $n_{1}$
and $n_{2}$ belong to two distinct fading blocks of $H_1(n)$, then the pattern
matrix column which separates $n_{1}$ and $n_{2}$ will have two
$1$ elements, i.e., one for $H_{1}(n)$ and the other for $H_{2}(n)$,
but this contradicts the BIA-feasible condition given by Lemma
1, which says each column should have a unique $1$ element. Thus $n_{1}$
and $n_{2}$ should belong to the same fading block of $H_{1}(n)$.
Repeating the same argument, we can prove that $n_{1}$, $n_{2}$,
$n_{3}$ and $n_{4}$ should fall in the same fading block of $H_{1}(n)$.
However, this will cause that the first row vector of the pattern
matrix has no $1$ element, and it contradicts the BIA-feasible
condition given by Lemma 1, which says each row should have a unique $1$
element. The contradiction completes the proof.
\end{IEEEproof}
Given $n_{\delta,2}$ and $n_{\delta,3}$, when $N$ is large, it
is formidable to determine whether BIA-feasible decomposition exists
by using the decomposition method shown in Fig. \ref{fig:BIA.implement.N4}.
To simplify the description
of the channel pattern for a homogeneous BC, we propose to cast the
channel fragment containing $0\leq n\leq4N-1$ into a 2-dimensional
pattern array. We define the 2-dimensional array as follows.
\begin{defn}
Divide the channel fragment containing $0\leq n\leq4N-1$ into $12$
groups, and number the groups from $0$ to $11$. During each group,
all channel states, i.e., $H_{i}(n)$, $i=1,2,3$, keep unchanged.
Denote $s_{i}$ as the size of the $i$th group, which is the number
of time slots in the group. The 2-dimensional pattern array is
formed by filling $s_{i}$ elements of $\pi_{i}$ along the $i$th
column. \end{defn}
\begin{example}
\label{ex: N4d1d2}To show how to form a pattern array, we use the
BIA-feasible BC described in Fig. \ref{fig:fading.ch.N4} as an example,
where $N=4$, $n_{\delta,2}=1$ and $n_{\delta,3}=2$. Starting from
$n=0$, the first group of $n_{\delta,2}-n_{\delta,1}=1$ symbols
have the constant channel coefficients. As shown in Fig. \ref{fig:ex.N4},
we fill the first column of the pattern array with $s_{0}=n_{\delta,2}-n_{\delta,1}=1$
elements of $\pi_{0}$. Then channel state variation happens at $n_{\delta,2}$,
but all channel coefficients keep unchanged over the group of $n_{\delta,2}\leq n\leq n_{\delta,3}-1$,
we then fill the second column of the pattern array with the same
number of symbols within the group ----- $s_{1}=n_{\delta,3}-n_{\delta,2}$
elements of $\pi_{1}$. Then, followed is a group of $N-n_{\delta,3}$
symbols with unchanged channel state, and we fill the third column
of the pattern array with $s_{2}=N-n_{\delta,3}$ elements of $\pi_{3}$.
The same construction process is repeated until the $11$th column
is filled. It is easy to see that $s_{0}+s_{1}+s_{2}=N$. Further
the same channel pattern repeats every $N$ symbols, we have $s_j=s_i$ if $i\equiv j\mod{3}$
for $u\in\{0,\cdots,11\}$, $j\in\{0,1,2\}$. Therefore, concerning the
pattern array, we have the following straightforward lemma.

\begin{figure}
\begin{centering}
\includegraphics[scale=1.1]{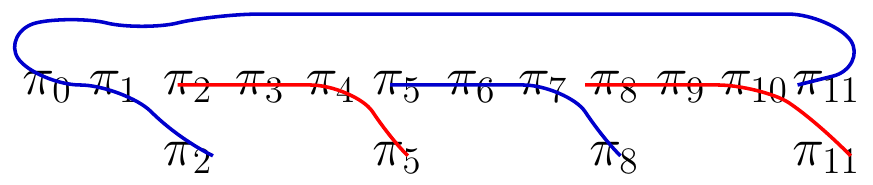}
\par\end{centering}

\caption{\label{fig:ex.N4}An example with $N=4$, $n_{\delta,2}=1$ and $n_{\delta,3}=2$.}

\end{figure}
\end{example}
\begin{lem}
\label{propty:si}Provided that a homogeneous BC with $n_{\delta,1}=0<n_{\delta,2}<n_{\delta,3}$,
the corresponding pattern array has
\begin{equation}
s_{i}=\begin{cases}
n_{\delta,2}, & i\equiv0\mod{3}\\
n_{\delta,3}-n_{\delta,2}, & i\equiv1\mod{3}\\
N-n_{\delta,3}, & i\equiv2\mod{3}
\end{cases}
\end{equation}
for $i\in\mathcal{Z}_{12}$ with $\mathcal{Z}_{12}$ representing
the integer ring on the base of $12$. If $n_{\delta,2}>n_{\delta,3}$, the roles of $n_{\delta,2}$ and
$n_{\delta,3}$ in the equation above should be exchanged.
\end{lem}

\subsection{BIA-feasibility in form of pattern array}

By using the pattern array model proposed above, the BIA-feasibility
problem of a BC can be easily formulated.
\begin{thm}
\label{thm:feasibility}Given a homogenous BC with the coherence time
$N$, the offsets $n_{\delta,1}=0$, $n_{\delta,2}>0$, $n_{\delta,3}>0$
and $n_{\delta,2}\neq n_{\delta,3}$, it is BIA-feasible if only if

(1), the corresponding pattern array can be completely decomposed
into $N$ 4-tuple $(\pi_{i},\pi_{i+1},\pi_{i+2},\pi_{i+3})$, where
$i,i+1,i+2,i+3\in\mathcal{Z}_{12}$.

or equivalently,

(2), the system of linear integer equations
\begin{equation}
s_{i}=\lambda_{i-3}+\lambda_{i-2}+\lambda_{i-1}+\lambda_{i},\quad i-3,i-2,i-1,i\in\mathcal{Z}_{12}\label{eq:feasible.lambda}
\end{equation}
i.e.,
\tiny
\begin{equation}
\left[\begin{array}{cccccccccccc}
1 & 0 &  &  &  & \cdots &  &  & 0 & 1 & 1 & 1\\
1 & 1 & 0 &  &  &  &  &  &  & 0 & 1 & 1\\
1 & 1 & 1 & 0 &  &  &  &  &  &  & 0 & 1\\
1 & 1 & 1 & 1 & 0 &  &  &  & \ddots &  &  & 0\\
0 & 1 & 1 & 1 & 1 & 0\\
 & 0 & 1 & 1 & 1 & 1 & 0\\
 &  & 0 & 1 & 1 & 1 & 1 & 0 &  &  &  & \vdots\\
 &  &  & 0 & 1 & 1 & 1 & 1 & 0\\
\vdots &  &  &  & 0 & 1 & 1 & 1 & 1 & 0\\
 &  & \ddots &  &  & 0 & 1 & 1 & 1 & 1 & 0\\
 &  &  &  &  &  & 0 & 1 & 1 & 1 & 1 & 0\\
0 &  &  & \cdots &  &  &  & 0 & 1 & 1 & 1 & 1
\end{array}\right]\left[\begin{array}{c}
\lambda_{0}\\
\lambda_{1}\\
\lambda_{2}\\
\lambda_{3}\\
\lambda_{4}\\
\lambda_{5}\\
\lambda_{6}\\
\lambda_{7}\\
\lambda_{8}\\
\lambda_{9}\\
\lambda_{10}\\
\lambda_{11}
\end{array}\right]=\begin{bmatrix}
s_{0}\\
s_{1}\\
s_{2}\\
s_{3}\\
s_{4}\\
s_{5}\\
s_{6}\\
s_{7}\\
s_{8}\\
s_{9}\\
s_{10}\\
s_{11}
\end{bmatrix}\label{eq:feasible.equations}
\end{equation}
\normalsize
\begin{equation}
\begin{bmatrix}
x & y & z\\
5 & 4 & 7
\end{bmatrix}
\end{equation}

has a solution $\{\lambda_{i}:i\in\mathcal{Z}_{12},\lambda_{i}\in\mathcal{N}_{0}=\{0,1,2,\cdots\}\}$
.\end{thm}
\begin{rem}
Before we start to prove the theorem, we elucidate the definition
of complete decomposition by referring to the example listed in Example
\ref{ex: N4d1d2}. In the pattern array of the example, the elements
$\pi_{i}$'s are connected by threads, each of which connects four
consecutive elements, i.e., a 4-tuple $(\pi_{i},\pi_{i+1},\pi_{i+2},\pi_{i+3})$.
Complete decomposition means that in the pattern array each element
$\pi_{i}$ is connected by one and only one thread. \end{rem}
\begin{IEEEproof}
Now we prove Theorem \ref{thm:feasibility} by beginning with the
part (1). According to Lemma \ref{lem:patternMatrix} and Lemma \ref{lem:no.delta.align},
the column generated by characterizing the transition from $\pi_{i}$
to $\pi_{i+1}$ for $\forall i\in\mathcal{Z}_{12}$ has only $1$
element since all $n_{\delta,i}$s are distinct. On the other hand,
according to Lemma \ref{propty:si}, for any 4-tuple $(\pi_{i},\pi_{i+1},\pi_{i+2},\pi_{i+3})$,
we have
\begin{equation}
N<\sum_{i}^{i+3}s_{i}<2N,
\end{equation}
which indicates each of $H_{1}(n)$, $H_{2}(n)$ and $H_{3}(n)$ undergoes
more than one coherence period but less than two coherent periods. This
means that in the pattern matrix formed by the 4-tuple, each row has
one and only one $1$ element. Therefore, by applying Lemma \ref{lem:patternMatrix},
any 4-tuple $(\pi_{i},\pi_{i+1},\pi_{i+2},\pi_{i+3})$ in the considered
homogeneous BC forms a feasible 4-symbol channel pattern. Clearly,
the considered homogeneous BC is feasible if the channel fragment
containing $4N$ consecutive symbols can be completely decomposed
into $N$ 4-tuple $(\pi_{i},\pi_{i+1},\pi_{i+2},\pi_{i+3})$ since
$4N$ is a period of the homogeneous BC. The necessary of the condition
can be proved by showing that it is not feasible if a 4-tuple $(\pi_{i_{1}},\pi_{i_{2}},\pi_{i_{3}},\pi_{i_{4}})\neq (\pi_{i},\pi_{i+1},\pi_{i+2},\pi_{i+3})$ for a $i\in\mathcal{Z}_{12}$.
The proof is relatively trivial, so we skip it for simplicity.

The equivalence between part (2) and part (1) is quite straightforward.
Let $\lambda_{i}\geq0$ be the number of 4-tuple $(\pi_{i},\pi_{i+1},\pi_{i+2},\pi_{i+3})$
in a complete decomposition. The group of $\pi_{i}$s is then fully
assigned to $\lambda_{i-3}$, $\lambda_{i-2}$, $\lambda_{i-1}$ and
$\lambda_{i}$. It is, then, clear that $\lambda_{i}$s should satisfy
the linear equations given by \eqref{eq:feasible.lambda}, which proves
part (2).
\end{IEEEproof}

\subsection{BIA-feasible region }

Previously, we show the BIA-feasible sufficient and necessary condition
in terms of the solvability of a system of linear equations. In this
part, we study the system of linear equations, and determine the region
of $s_{i}$ such that the system is solvable. Such a region of $s_{i}$
is called as feasible region. Further, we transfer the feasible region
into the one represented by $n_{\delta,i}$.
\begin{thm}
\label{thm:weak.condition}If a pattern array is BIA-feasible, i.e.,
the system of linear equations \eqref{eq:feasible.equations} has
solutions, then it must be satisfied that $\max(s_{0},s_{1},s_{2})\leq2\min(s_{0},s_{1},s_{2})$. \end{thm}
\begin{IEEEproof}
Without loss of generality, we let $s_{0}\leq s_{1}\leq s_{2}$. We
prove the theorem by contradiction. Assume that $s_{2}>2s_{0}$, and
the corresponding linear system has solutions. Since the linear system
is solvable, it must have
\begin{eqnarray}
s_{0} & = & \lambda_{9}+\lambda_{10}+\lambda_{11}+\lambda_{0}\geq\lambda_{11}+\lambda_{0}\\
s_{3} & = & \lambda_{0}+\lambda_{1}+\lambda_{2}+\lambda_{3}\geq\lambda_{1}+\lambda_{2}
\end{eqnarray}
Using $s_{3}=s_{0}$ from Lemma \ref{propty:si}, we have
\begin{equation}
s_{2}=\lambda_{11}+\lambda_{0}+\lambda_{1}+\lambda_{2}\leq s_{0}+s_{3}=2s_{0}
\end{equation}
which contradicts with the assumption $s_{2}>2s_{0}$. The proof
is complete.\end{IEEEproof}
\begin{thm}
\label{thm:strong.codition}If a pattern array is BIA-feasible, i.e.,
the system of linear equations \eqref{eq:feasible.equations} has
solutions, then it must be satisfied that

(1), if $s_{k}\leq s_{j}\leq s_{i}$ with $(k,j,i)$ be a permutation
of $(0,1,2)$, then $\exists x,y\in\mathcal{N}_{0}$ such that
\begin{subequations}
\begin{eqnarray}
x+y & = & s_{k}\\
s_{j}-x & \leq & s_{k}\\
s_{i}-y & \leq & s_{k}
\end{eqnarray}
\end{subequations}
or equivalently,

(2), $\sum_{i=0}^{2}s_{i}\leq4\min(s_{0},s_{1},s_{2})$.\end{thm}
\begin{IEEEproof}
We start with the proof of part (1). It is easy to see that in
order to prove part (1), we only need to prove that if $s_{i}+s_{j}>3s_{k}$,
then the pattern array is not BIA-feasible. Without loss of generality,
we let $s_{0}\leq s_{1}\leq s_{2}$. Now, given $s_{1}+s_{2}>3s_{0}$,
we assume the linear system has a solution $\{\lambda_{i}:i\in\mathcal{Z}_{12},\lambda_{i}\in\mathcal{N}_{0}\}$.

First, from $\lambda_{9}+\lambda_{10}+\lambda_{11}+\lambda_{0}=s_{0},$
we get $\lambda_{10}+\lambda_{11}+\lambda_{0}=s_{0}-\lambda_{9}\leq s_{0}$.
Substituting this inequality into $\lambda_{1}=s_{1}-(\lambda_{10}+\lambda_{11}+\lambda_{0})$
gives the inequality
\begin{equation}
\lambda_{1}\geq s_{1}-s_{0}.
\end{equation}
Then substituting it into $\lambda_{2}+\lambda_{3}+\lambda_{4}=s_{4}-\lambda_{1}$,
in conjunction with $s_{4}=s_{1}$, we get
\begin{equation}
\lambda_{2}+\lambda_{3}+\lambda_{4}\leq s_{0}.
\end{equation}
Further applying the new inequality to $\lambda_{5}=s_{5}-(\lambda_{2}+\lambda_{3}+\lambda_{4})$,
along with $s_{5}=s_{2}$, we have
\begin{equation}
\lambda_{5}\geq s_{2}-s_{0}.
\end{equation}

Secondly, from $s_{3}=\lambda_{0}+\lambda_{1}+\lambda_{2}+\lambda_{3}$
and $s_{3}=s_{0}$, we can get $\lambda_{1}+\lambda_{2}+\lambda_{3}\leq s_{0}$.
Applying it and $s_{4}=s_{1}$ to $\lambda_{4}=s_{4}-(\lambda_{1}+\lambda_{2}+\lambda_{3})$,
we have
\begin{equation}
\lambda_{4}\geq s_{1}-s_{0}.
\end{equation}

Adding these two inequalities about $\lambda_{5}$ and $\lambda_{4}$,
along with the assumption $s_{1}+s_{2}>3s_{0}$, shows that
\begin{equation}
\lambda_{4}+\lambda_{5}\geq s_{1}+s_{2}-2s_{0}>s_{0}.
\end{equation}
However, this contradicts with the fact that $\lambda_{3}+\lambda_{4}+\lambda_{5}+\lambda_{6}=s_{6}=s_{0}$.
So the proof for the part (1) is complete.

Now we prove the equivalence between part (1) and part (2).
On one hand, given the premise of part (1), adding the three formula
in part (1) would lead to the condition of part (2). On the
other hand, provided that $s_{0}\leq s_{1}\leq s_{2}$, the condition
of part (2) $\sum_{i=0}^{2}s_{i}\leq4\min(s_{0},s_{1},s_{2})$
gives $s_{1}+s_{2}\leq3s_{0}$. Now let $x=s_{1}-s_{0}$, $y=s_{0}-x$,
they will satisfy the condition given in the part (1). So, the equivalence
is proved.\end{IEEEproof}
\begin{rem}
In fact, the necessary condition given by Theorem \ref{thm:strong.codition}
is stronger than the condition given by Theorem \ref{thm:weak.condition}.
To see this, supposing $s_{0}\leq s_{1}\leq s_{2}$, we only need to
show that $\sum_{i=0}^{2}s_{i}\leq4s_{0}$ leads to $s_{2}\leq2s_{0}$.
It is clear that $\sum_{i=0}^{2}s_{i}\leq4s_{0}$ gives $s_{1}+s_{2}\leq3s_{0}$.
Combining this inequality with $s_{1}\geq s_{0}$ reaches $s_{2}\leq2s_{0}$.
Nevertheless, Theorem \ref{thm:weak.condition} shows a simple criterion
to determine a pattern array is not feasible.
\end{rem}
In the following we prove that the condition given by Theorem~\ref{thm:strong.codition}
is also a sufficient condition.

\begin{figure}
\begin{centering}
\subfloat[\label{fig:implement.A}The BIA implementation with $\lambda_{2}=3$.]{\begin{centering}
\includegraphics[scale=0.6]{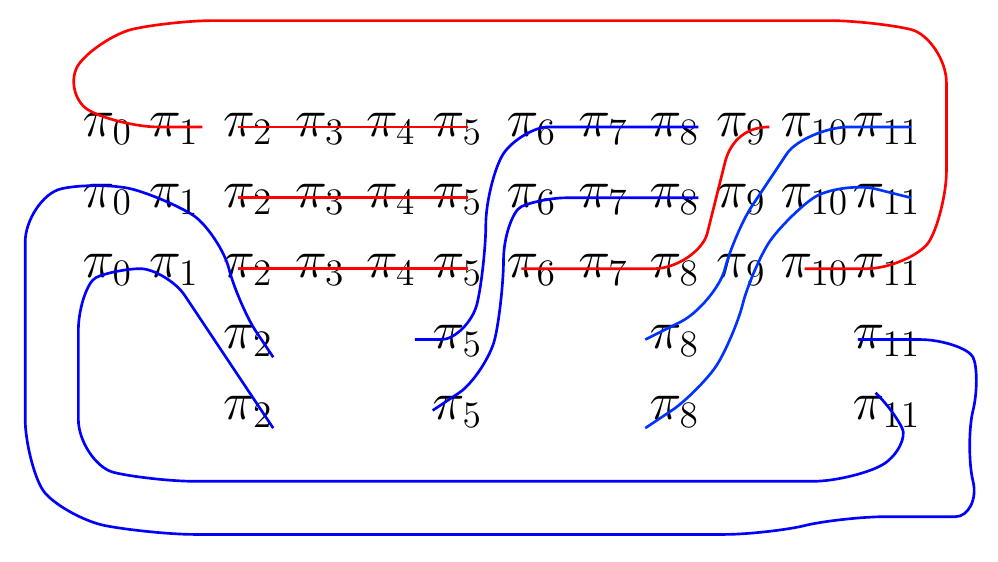}
\par\end{centering}
}\subfloat[\label{fig:implement.B}The BIA implementation with $\lambda_{2}=2$.]{\begin{centering}
\includegraphics[scale=0.6]{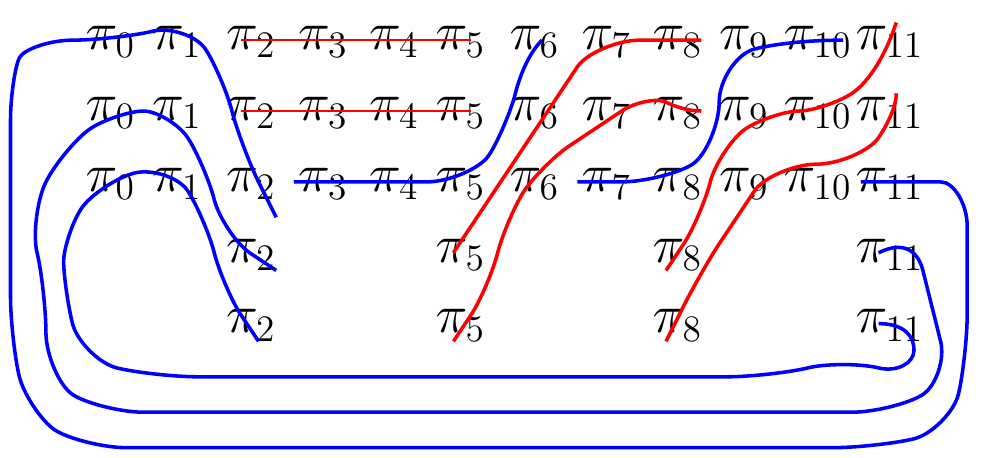}
\par\end{centering}
}
\par\end{centering}
\caption{\label{fig:two.implement}Different BIA implementation for a homogeneous
BC with $N=13$.}
\end{figure}

\begin{thm}
\label{thm:sufficient.condition}If a pattern array satisfies $\sum_{i=0}^{2}s_{i}\leq4\min(s_{0},s_{1},s_{2})$,
then it is BIA-feasible. \end{thm}
\begin{IEEEproof}
Without loss of generality, we let $s_{0}\leq s_{1}\leq s_{2}$.
Since $\sum_{i=0}^{2}s_{i}\leq4\min(s_{0},s_{1},s_{2})$, we can easily
get $s_{1}+s_{2}\leq3s_{0}$. Let $x=s_{1}-s_{0}$, and $y=s_{0}-x$.
Clearly, $y\geq0$ since $s_{1}\leq2s_{0}$ according to Theorem \ref{thm:weak.condition},
and thus $x\leq s_{0}$. And also this assignment of $x$ and $y$
satisfies the condition $s_{2}-y=s_{2}+s_{1}-2s_{0}\leq s_{0}$. Given
this pair of $x$ and $y$, we can easily prove that a feasible solution
for the linear system shown in \eqref{eq:feasible.equations} is
\begin{equation}
\left[\begin{array}{c}
\lambda_{0}\\
\lambda_{1}\\
\lambda_{2}\\
\lambda_{3}\\
\lambda_{4}\\
\lambda_{5}\\
\lambda_{6}\\
\lambda_{7}\\
\lambda_{8}\\
\lambda_{9}\\
\lambda_{10}\\
\lambda_{11}
\end{array}\right]=\left[\begin{array}{c}
0\\
x\\
y\\
0\\
s_{1}-s_{0}\\
s_{2}-s_{1}+x\\
3s_{0}-s_{1}-s_{2}\\
s_{1}-s_{0}\\
s_{2}-s_{1}+x\\
0\\
2s_{0}-s_{2}\\
s_{2}-s_{0}
\end{array}\right].
\end{equation}
Therefore, the corresponding pattern array is BIA-feasible, which
proves the theorem.
\end{IEEEproof}

\begin{figure}
\begin{centering}
\subfloat[BIA-feasible region for $N=20$.]{\begin{centering}
\includegraphics[scale=0.62]{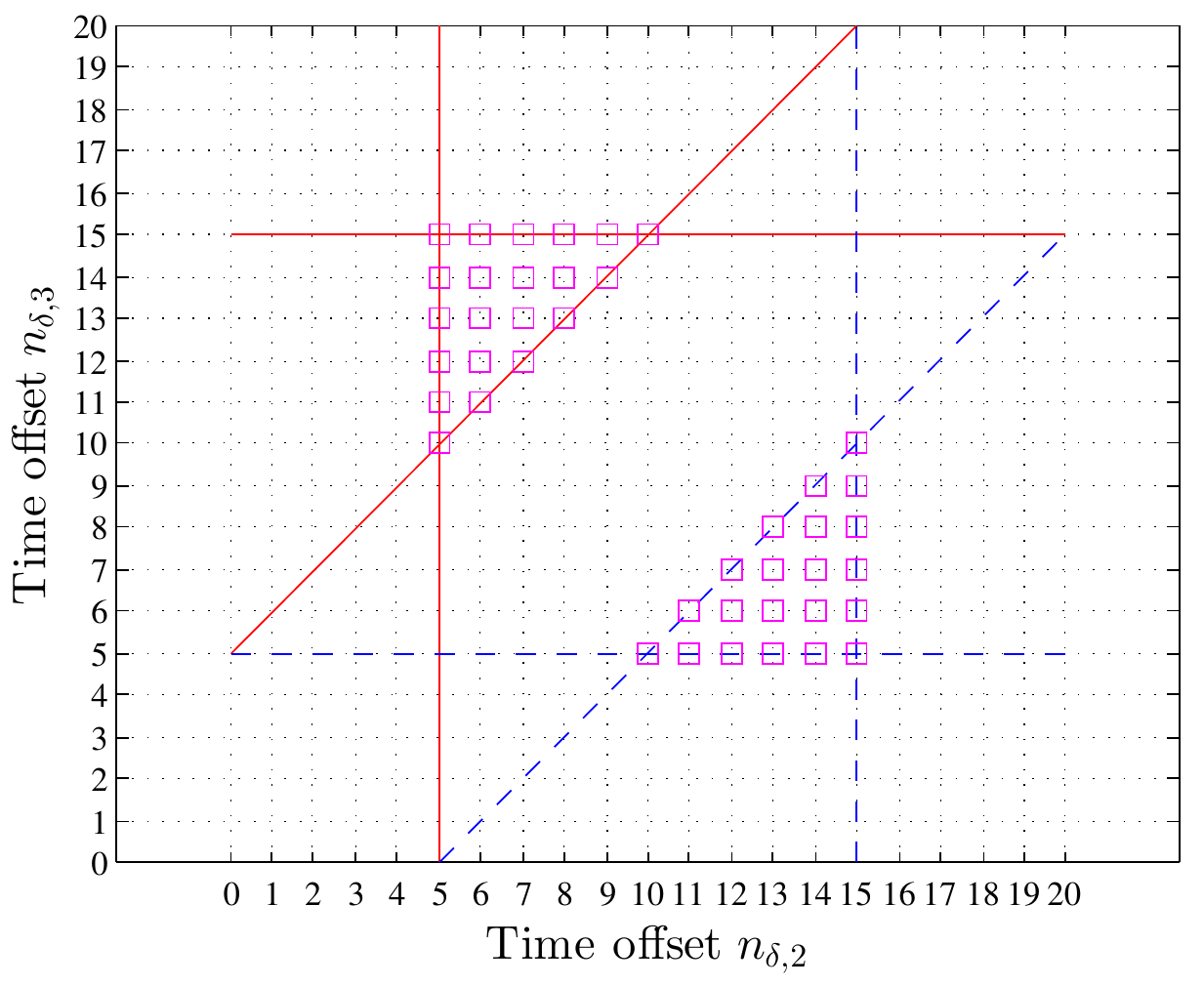}
\par\end{centering}

}\subfloat[BIA-feasible region for $N=21$.]{\begin{centering}
\includegraphics[scale=0.62]{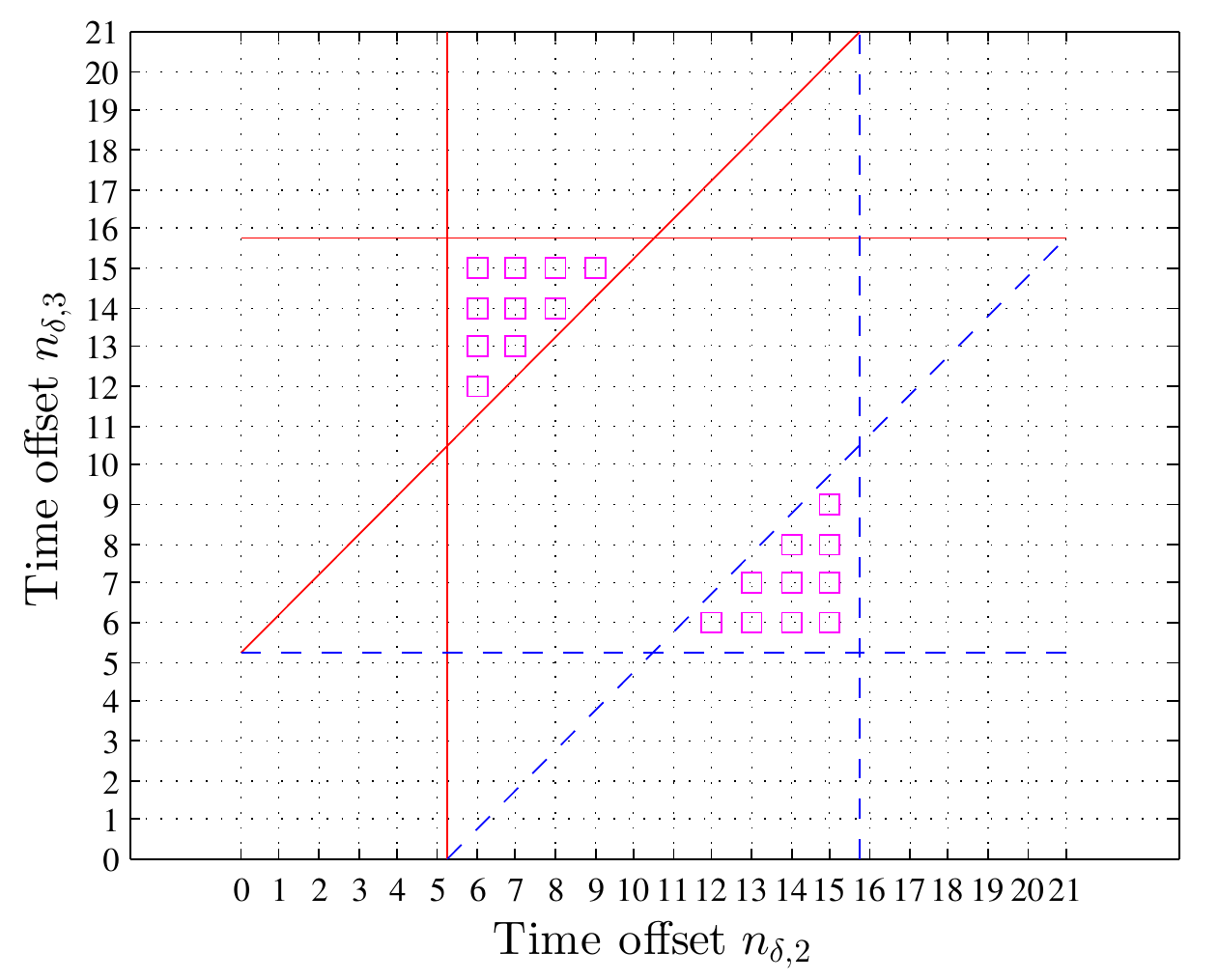}
\par\end{centering}

}
\par\end{centering}

\caption{\label{fig:feasible.reg.N21}The BIA-feasible regions in terms of
$n_{\delta,2}$ and $n_{\delta,3}$ for $N=20$ and $N=21$, respectively. Box marker
represents BIA-feasible point. }
\end{figure}

\begin{rem}
It is worthy to point out that the feasible solution given in Theorem
\ref{thm:sufficient.condition} is not unique. As shown in Fig. \ref{fig:two.implement},
two different solutions are feasible for the homogeneous BC with $N=13$,
$n_{\delta,2}=3$ and $n_{\delta,3}=6$. In Fig. \ref{fig:implement.A},
the number of threads starting from $\pi_{2}$ is $\lambda_{2}=3$,
while in the Fig. \ref{fig:implement.B} this number is $\lambda_{2}=2$.
\end{rem}
By applying Theorem \ref{thm:sufficient.condition} into Lemma \ref{propty:si}, we can get the BIA-feasible region characterized
by the parameter $N$ and $n_{\delta,i}$s, which gives the following
theorem.
\begin{thm}
\label{thm:feasible.region.n}Given $n_{\delta,1}=0,$ the homogeneous 3-user
BC is BIA-feasible if the 2-tuple $(n_{\delta,2},n_{\delta,3})$ satisfies
one of the following conditions

(1), when $n_{\delta,2}<n_{\delta,3}$, then
\begin{subequations}
\begin{eqnarray}
n_{\delta,3} & \leq & 3N/4\\
n_{\delta,2} & \geq & N/4\\
n_{\delta,3}-n_{\delta,2} & \geq & N/4
\end{eqnarray}

\end{subequations}

(2), when $n_{\delta,2}>n_{\delta,3}$, then the conditions are given
by the same group of inequalities as above except that the roles of
$n_{\delta,2}$ and $n_{\delta,3}$ should be exchanged. \end{thm}
\begin{IEEEproof}
Suppose $n_{\delta,2}<n_{\delta,3},$ then from Lemma \ref{propty:si}
we have $s_{0}=n_{\delta,2}$, $s_{1}=n_{\delta,3}-n_{\delta,2}$
and $s_{2}=N-n_{\delta,3}$. From $\sum_{i=0}^{2}s_{i}\leq4\min(s_{0},s_{1},s_{2})$,
along with $\sum_{i=0}^{2}s_{i}=N$, we have $4\min(s_{0},s_{1},s_{2})\geq N$,
therefore
\begin{equation}
\min(s_{0},s_{1},s_{2})\geq N/4.
\end{equation}
By using $s_{i}\geq\min(s_{0},s_{1},s_{2})$, together with the definition
of $s_{i}$ in terms of $n_{\delta,2}$ and $n_{\delta,3}$, we can
get
\begin{eqnarray}
n_{\delta,2} & = & s_{0}\geq N/4,\\
n_{\delta,3}-n_{\delta,2} & = & s_{1}\geq N/4,\\
N-n_{\delta,3} & = & s_{2}\geq N/4,
\end{eqnarray}
which together prove the theorem.
\end{IEEEproof}
As an example, Fig. \ref{fig:feasible.reg.N21} shows the BIA-feasible
regions for $N=20$ and $N=21$, respectively. The feasible region is achieved by
using the feasible conditions given by Theorem \ref{thm:feasible.region.n}. As shown in
the figure, for the case of $N=20$ there are $42$ feasible points
out of the total $20^{2}=400$ points, resulting in the feasible ratio
$42/400=0.105$; for the case of $N=21$ there are $20$ feasible
points out of the total $21^{2}=441$, and thus the feasible ratio
is $20/441=0.0454$. This comparison shows that the feasible ratio
is not a monotonic increasing function of coherence time $N$.

\subsection{Probability of finding feasible $3$-user $2\times1$ BC from $K\geq4$
users}

Previously, when there are $3$ users, we studied the sufficient and
necessary BIA-feasible condition on 2-tuple $(n_{\delta,2},n_{\delta,3})$
provided $n_{\delta,1}=0$. In this part, we examine the probability
that among $K$ users there exists a 3-tuple $(n_{\delta,i},n_{\delta,j},n_{\delta,k})$
forming a BIA-feasible 3-user $2\times1$ homogenous BC.

We first give a lemma showing the BIA-feasible condtion on $(n_{\delta,1},n_{\delta,2},n_{\delta,3})$
without the assumption of $n_{\delta,1}=0$.
\begin{lem}
\label{lem:3.user.no.benchmark}3 users form a BIA-feasible BC if
$|n_{\delta,i}-n_{\delta,j}|\geq\left\lceil \frac{N}{4}\right\rceil $
holds for any pair of $i\neq j$.
\end{lem}
\begin{figure}
\begin{centering}
\includegraphics[scale=1.5]{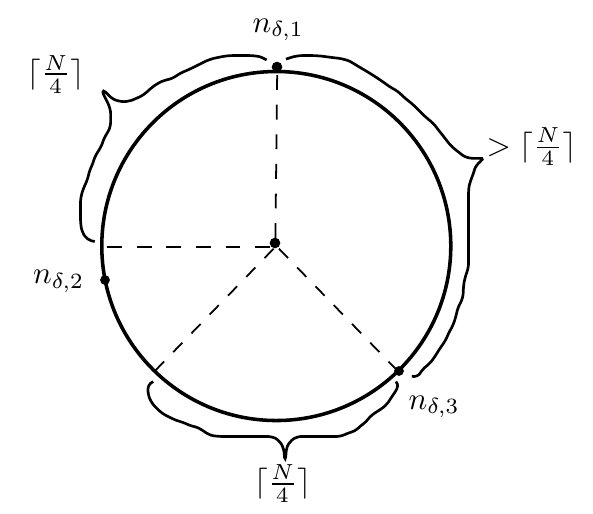}
\par\end{centering}

\caption{\label{fig:3.user.ring}A BIA-feasible 3-tuple $(n_{\delta,1},n_{\delta,2},n_{\delta,3})$.}
\end{figure}

\begin{IEEEproof}
Illustratively, we can visualize the condition by Fig. \ref{fig:3.user.ring},
in which any pair of $n_{\delta,i}$ and $n_{\delta,j}$ is separated
by at least $\lceil\frac{N}{4}\rceil$. The lemma can be easily proved
by setting one user as the benchmark, say $n_{\delta,1}=0$, and then
applying Theorem \ref{thm:feasible.region.n}.
\end{IEEEproof}
Next we show another lemma which is about to be used in the following analysis.
\begin{lem}
Suppose there are $n$ labeled boxes, and $\Theta$ labeled balls.
Given $\mu \leq \min\{n,\Theta\}$ boxes, the number of ways to put the balls into the boxes
such that the $\mu$ boxes are not empty is given by
\begin{equation}
\gamma(n,\Theta,\mu)=\sum_{k=\mu}^{\Theta}\binom{\Theta}{k}\mu!S(k,\mu)(n-\mu)^{\Theta-k},
\end{equation}
where $S(k,\mu)=\frac{1}{\mu!}\sum_{j=0}^{\mu}(-1)^{\mu-j}\binom{\mu}{j}j^{k}$
is the Stirling number of the second kind \cite{Stanley2011}. \end{lem}
\begin{IEEEproof}
We divide the ball assignment process into two steps. Firstly we randomly
choose $k\geq\mu$ balls, which has $\binom{\Theta}{k}$ ways, and
put the chosen balls into the $\mu$ boxes such that each box has
at least one balls, which has $\mu!S(k,\mu)$ ways. Secondly we randomly
put the rest $\Theta-k$ balls into the rest $n-\mu$ boxes. To combine
these two steps and sum over $\mu\leq k\leq\Theta$ proves the lemma.
\end{IEEEproof}
When there are $K$ users, by using the two lemmas above, we can count
the events in which no three users' offsets can form the feasible
ring as shown in Fig. \ref{fig:3.user.ring}. To ease the derivation,
we assume $N$ is a multiplicity of $4$, that is, $\frac{N}{4}\in\mathcal{N}=\{1,2,\cdots\}$.

\begin{figure}
\begin{raggedright}
\subfloat[\label{fig:Type-1}Type 1.]{\begin{centering}
\includegraphics[scale=1.5]{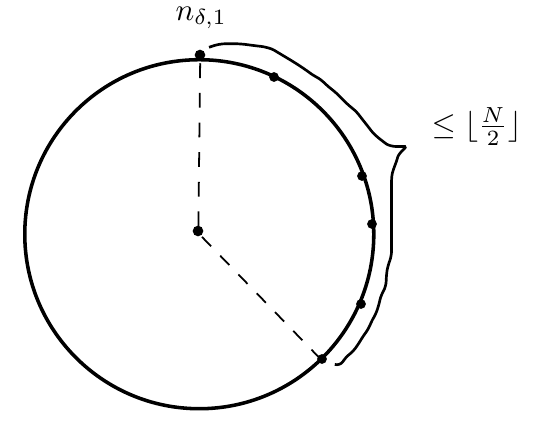}
\par\end{centering}

}\subfloat[\label{fig:Type-2}Type 2]{\centering{}
\includegraphics[scale=1.5]{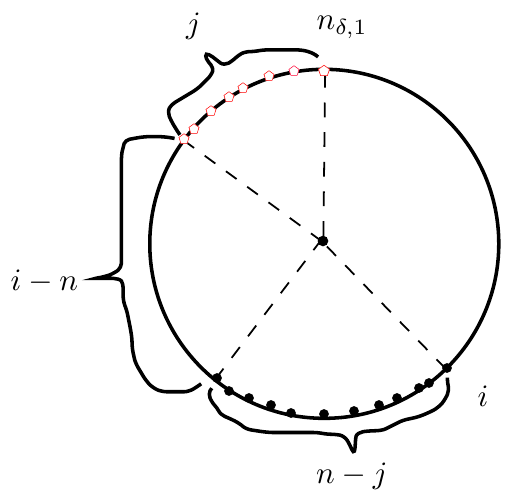}}
\par\end{raggedright}

\caption{\label{fig:thm.ball.box}Events with no BIA-feasible 3-tuple.}
\end{figure}

\begin{figure}
\begin{centering}
\includegraphics[scale=.8]{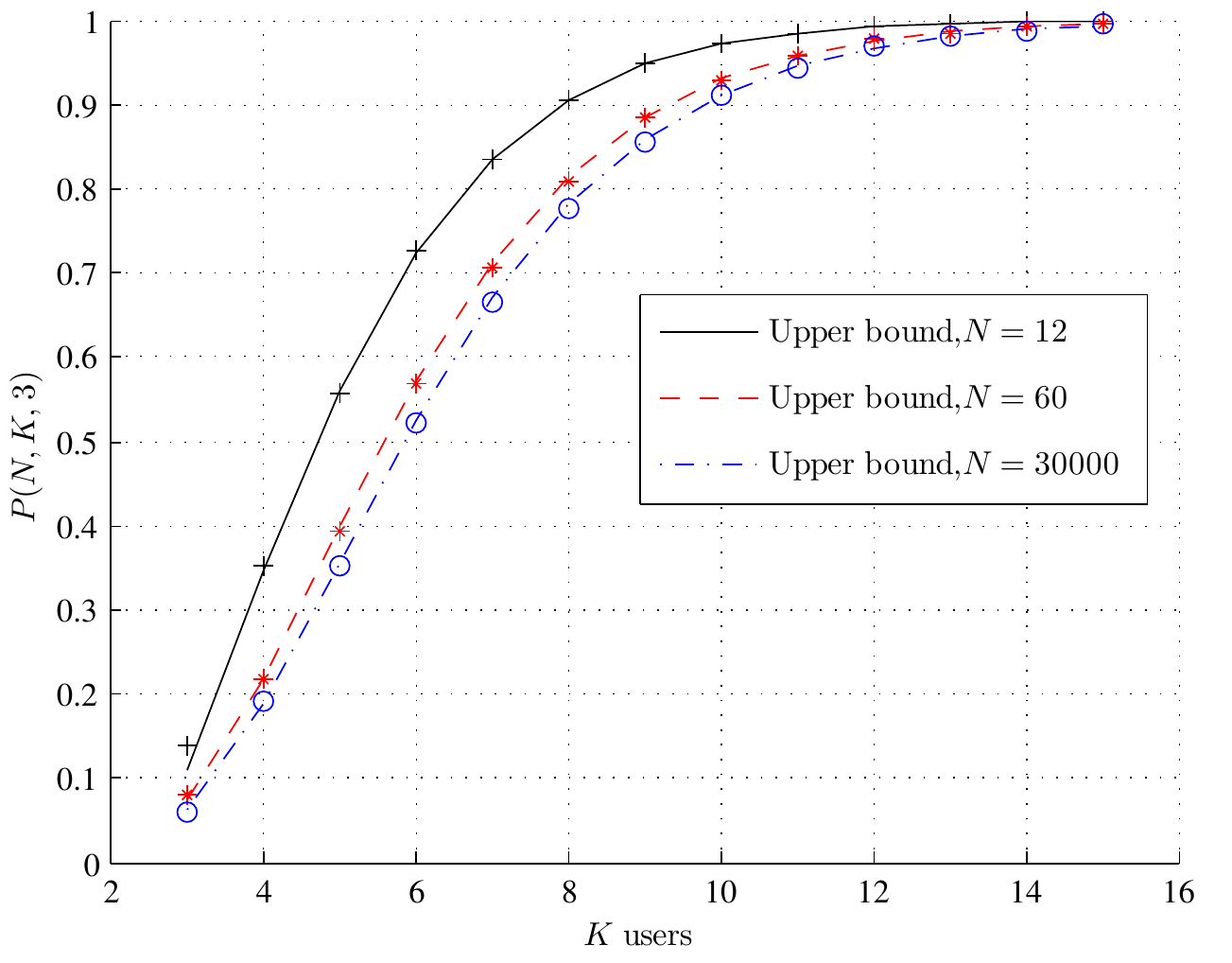}
\par\end{centering}
\caption{\label{fig:simVsbound}$P_{up}(N,K,3)$ vs simulation results. Simulation
results are shown by marks.}
\end{figure}

\begin{thm}
Given $N$ the coherence time subject to $\frac{N}{4}\in\mathcal{N}=\{1,2,\cdots\}$,
and $K\geq4$ the number of users, let $f(N,K,3)$ be the number of events in which no three users' offsets can meet the condition given by Lemma~\ref{lem:3.user.no.benchmark} and form a 3-user BIA-feasible ring as shown in
Fig. \ref{fig:3.user.ring}, then
\begin{eqnarray}
f(N,K,3) & > & 1+\sum_{n=2}^{N/2}\left(2[n^{K-1}-(n-1)^{k-1}]+(n-2)[n^{K-1}-2(n-1)^{K-1}+(n-2)^{K-1}]\right)\nonumber \\
 &  & + (2^{K-1}-1)+\sum_{n=3}^{N/4+1}(n-1)\left[2(n-3)\gamma(n,K-1,3)+3\gamma(n,K-1,2)\right]\nonumber \\
 &  & +\sum_{n=3}^{N/4+1}(n-1)(n-3)\left[\frac{1}{2}(n-4)\gamma(n,K-1,4)+\gamma(n,K-1,3)\right] + \nonumber \\
 &  & \sum_{n=N/4+2}^{N/2}(\frac{N}{2}-n+1)(n-1)\left(2\gamma(n,K-1,3)+\frac{1}{2}(n-4)\gamma(n,K-1,4)\right)\\
 &=&
 f_{low}(N,K,3)
 \label{eq:no.BIA.events.num}
\end{eqnarray}
where

\begin{subequations}

\begin{eqnarray}
\gamma(n,K-1,2) & = & n^{K-1}-2(n-1)^{K-1}+(n-2)^{K-1},\\
\gamma(n,K-1,3) & = & n^{K-1}-3(n-1)^{K-1}+3(n-2)^{K-1}-(n-3)^{K-1},\\
\gamma(n,K-1,3) & = & n^{K-1}-4(n-1)^{K-1}+6(n-2)^{K-1}-4(n-3)^{K-1}+(n-4)^{K-1}.
\end{eqnarray}

\end{subequations}\end{thm}
\begin{IEEEproof}
We cast this problem into a ball-box problem, in which $N$ labeled
boxes form a ring, and a user is denoted by a ball, the user's offset
is denoted by the label of the box which contains the ball. As illustrated
in Fig. \ref{fig:thm.ball.box}, we set $n_{\delta,1}$ as the benchmark,
and label the box containing it as Number 1. We divide all events
meeting the no-BIA condition into two types. The first is that the
number of the boxes in the arc which contains all users is no greater
than $\frac{N}{2}$, and the second that the arc is larger than $\frac{N}{2}$.
In the following, we refer to the length of an arc as the number of
the boxes in the arc.

\textbf{Type I}), We start with the first type. Given an arc with the
length $n\leq\frac{N}{2}$ as shown in Fig. \ref{fig:Type-1}, we
count the event number by applying the similar argument developed
in \cite{Zhou2012}. If $n_{\delta,1}$ is one end point of the arc,
then the rest $K-1$ users can be randomly loaded into the $n$ boxes
on the arc subject to the condition that the other end of the arc
must be occupied by at least one user, resulting in the number of
such events $2[n^{K-1}-(n-1)^{K-1}]$, where $2$ reflects $n_{\delta,1}$
can be either of the two end points. If $n_{\delta,1}$ is not any
end point of the arc, then the position of the arc relative to $n_{\delta,1}$
has $n-2$ possibilities, and for each possibility the rest $K-1$
users can be randomly loaded but the two end points of the arc must be
occupied, resulting in the number of such events
$(n-2)[n^{K-1}-2(n-1)^{K-1}+(n-2)^{K-1}]$, i.e., $(n-2)\gamma(n,K-1,2)$. Combining them, and summing
over $1\leq n\leq\frac{N}{2}$, we get the number of events for the
type I
\begin{equation}
f_{1}(N,K,3)=1+\sum_{n=2}^{N/2}\left(2[n^{K-1}-(n-1)^{K-1}]+
(n-2)\gamma(n,K-1,2)\right).
\end{equation}

\textbf{Type II}), Now we count the type II events. As shown in Fig.
\ref{fig:Type-2}, we number the boxes counter-clockwise by increasing
integer. Assume all users are located in two arcs, which occupy $n$
positions in total; one arc is filled by users denoted by pentagons,
the other by users denoted by dots.

(II,a): If $n_{\delta,1}$ is one end point of the pentagon arc, the
position of the end point of the dot arc $i$ should satisfy the condition
$\frac{N}{2}+1\leq i\leq n+\frac{N}{2}-1$, otherwise the event would
belong to the type I above. Given such a $i$, \emph{we count the events with
$j\leq\frac{N}{4}$ and $n-j\leq\frac{N}{4}$}.

Given $n=2$, then $i=\tfrac{N}{2}+1$, the rest $K-1$ balls are randomly put
into two boxes subject to that the other box must be occupied, resulting in the number of such events $2^{K-1}-1$.

Given $3\leq n\leq\frac{N}{4}+1$ and $i$, when the pentagon arc
length $j$ satisfies $2\leq j\leq n-2$, the user $n_{\delta,1}$
can be either of the end points of the arc. So the number of possibilities
of the arc combination with $n_{\delta,1}$ being one end point is
$2(n-3)$, and the rest $K-1$ users can be randomly located at the
two arcs with $n$ positions subject to that the other three
end points of the two arcs must be occupied, resulting
in the number of such events $2(n-3)\gamma(n,K-1,3)$. When $j=1$,
the two ends of the pentagon arc is the same, and thus the two arcs
are determined by three users, consequentially there are $K-1$ users
to be located subject to that the two end points of the dot arc
must be occupied, resulting in the number of events $\gamma(n,K-1,2)$. When $j=n-1$,
the dot arc has only one end point, resulting in the number of event
$2\gamma(n,K-1,2)$, in which $2$ reflects $n_{\delta,1}$ can be
either of the two end points of the pentagon arc. Combining $j=1$,
$2\leq j\leq n-2$, and $j=n-1$, we get $2(n-3)\gamma(n,K-1,3)+3\gamma(n,K-1,2)$
the number of such events, in which $n_{\delta,1}$is an end point,
given $3\leq n\leq\frac{N}{4}+1$ and $i$.

Given $\frac{N}{4}+2\leq n\leq\frac{N}{2}$ and $i$, the pentagon
arc length can not be $j=1$ or $j=n-1$, but $n-\frac{N}{4}\le j\leq\frac{N}{4}$.
Referring to the argument above for $j\neq1$ and $j\neq n-1$, we
get the number of even $2(\frac{N}{2}-n+1)\gamma(n,K-1,3)$, where
$\frac{N}{2}-n+1$ stands for the possible choices of $j$ subject to $n-\frac{N}{4}\le j\leq\frac{N}{4}$.

(II.b): If $n_{\delta,1}$ is not one end point of the pentagon arc,
but an internal point of the arc, which only happens when $j\geq3$.
Then, when $4\leq n\leq\frac{N}{4}+1$ and $3\leq j\leq n-2$, the
user $n_{\delta,1}$ can take $j-2$ internal positions, resulting
in $\sum_{j=3}^{n-2}(j-2)$ possibilities. The rest $K-1$ users can
be randomly located subject to that each of the end points of the
two arcs must be occupied by at least one user, resulting in $\sum_{j=3}^{n-2}(j-2)\gamma(n,K-1,4)$
possibilities of assignment. When $j=n-1$, the dot arc only has one
end point, and the three end points must be occupied, generating $(j-2)\gamma(n,K-1,3)=(n-3)\gamma(n,K-1,3)$
possible assignments. Combining $3\leq j\leq n-2$ and $j=n-1$, we
get the number of events $\sum_{j=3}^{n-2}(j-2)\gamma(n,K-1,4)+(n-3)\gamma(n,K-1,3)$,
in which $n_{\delta,1}$ is located in an internal position, given
$4\leq n\leq\frac{N}{4}+1$ and $i$.

If $n_{\delta,1}$ is not one end point, and $\frac{N}{4}+2\leq n\leq\frac{N}{2}$,
the length of pentagon arc is conditioned on $n-\frac{N}{4}\le j\leq\frac{N}{4}$,
which excludes the chance of $j=n-1$. Then referring to the argument
above for $j\neq n-1$, we get the number of events $\sum_{j=n-N/4}^{N/4}(j-2)\gamma(n,K-1,4)$.

In summary, combining these two subcases for type II, and summing
over $\frac{N}{2}+1\leq i\leq n+\frac{N}{2}-2$ and over either $2\leq n\leq\frac{N}{4}+1$
or $\frac{N}{4}+2\leq n\leq\frac{N}{2}$, we get the number of events
for type II
\begin{eqnarray}
f_{2}(N,K,3) & = & (2^{K-1}-1)+\sum_{n=3}^{N/4+1}\sum_{i=N/2+1}^{n+N/2-1}\left[2(n-3)\gamma(n,K-1,3)+3\gamma(n,K-1,2)\right]\nonumber \\
 &  & +\sum_{n=4}^{N/4+1}\sum_{i=N/2+1}^{n+N/2-1}\left[\sum_{j=3}^{n-2}(j-2)\gamma(n,K-1,4)+(n-3)\gamma(n,K-1,3)\right] + \nonumber \\
 &  & \sum_{n=N/4+2}^{N/2}\sum_{i=N/2+1}^{n+N/2-1}\left(2(\frac{N}{2}-n+1)\gamma(n,K-1,3)+\sum_{j=n-N/4}^{N/4}(j-2)\gamma(n,K-1,4)\right)\nonumber \\
\end{eqnarray}

Finally, the number of events which contains no BIA-feasible 3-tuple
is given by adding $f_{1}(N,K,3)$ and $f_{2}(N,K,3)$, which proves the
theorem.

The above event counting is only a lower bound of $f(N,K,3)$ because there are
some events which are against the constraints of $j\leq\frac{N}{4}$ and 
$n-j\leq\frac{N}{4}$, but also generate no BIA-feasible 3-tuple. For instance, 
they include the event that $K=5$ balls are equally separated. 
\end{IEEEproof}
\begin{cor}
The equation \eqref{eq:no.BIA.events.num} also holds for $K=3$. \end{cor}
\begin{IEEEproof}
This result can be easily verified by realizing that $\gamma(n,2,3)=\gamma(n,2,4)=0$.
\end{IEEEproof}
It is physically justified that $n_{\delta,i}$ is uniformly distributed
over $[0,N-1]$ \cite{Zhou2012}. Based on this uniform distribution
assumption, we derive the probability of finding a $3$-user $2\times1$
homogeneous BC from $K$ users in the following theorem.
\begin{thm}
Given the $2\times1$ BC network with $K$ homogeneous users, let $P(N,K,3)$ be
the probability that the transmitter finds three users among the $K$
users to form a BIA-feasible $3$-user $2\times1$ MISO BC. Then
\begin{equation}\label{eq:PNK3}
P(N,K,3)=1-\frac{f(N,K,3)}{N^{K-1}}\leq 1-\frac{f(N,K,3)}{N^{K-1}}=P_{up}(N,K,3).
\end{equation}
\end{thm}
\begin{IEEEproof}
The result is clear since distributing $n_{\delta,k}$, $2\leq k\leq K$
over $\mathcal{Z}_{N}$ uniformly is equivalent to casting $K-1$
labeled balls uniformly into the ring of $N$ labeled boxes.
\end{IEEEproof}
\begin{rem}
Fig.~\ref{fig:simVsbound} shows that the derived upper bound of $P(N,K,3)$ is 
quite tight. 
\end{rem}

\section{BIA-feasibility for $K$-user $2\times1$ homogeneous BC}

Previously we derived the sufficient and necessary BIA-feasible condition
on offsets for the homogeneous $3$-user $2\times1$ BC. In this section,
we will extend the investigation to any homogeneous $K$-user $2\times1$
BC with $K\geq2$.
\begin{thm}
For the general $K$-user $2\times1$ homogeneous BC with $K\geq2$,
the sufficient and necessary BIA-feasible condition is
\begin{equation}
\sum_{k=0}^{K-1}s_{k}\leq(K+1)\min\{s_{k}:k\in\mathcal{Z}_{K}\}. \label{eq:K.user.SN.cond}
\end{equation}
 \end{thm}
\begin{IEEEproof}
\textbf{Necessary condition}: We start with proving the condition
is necessary. Referring to Theorem \ref{thm:feasibility}, with similar proof, we can show
that to the BIA-feasibility of a $K$-user $2\times1$ BC channel is equivalent to the solvability
of the following linear system
\begin{equation}
s_{i}=\sum_{j=i-K,j\in\mathcal{Z}_{K(K+1)}}^{i}\lambda_{j},\quad i\in\mathcal{Z}_{K(K+1)},\lambda_{j}\in\mathcal{N}_{0},\label{eq:K.user.lambda}
\end{equation}
subject to
\begin{eqnarray}
\sum_{i=0}^{K-1}s_{i} & = & N\\
s_{i} & = & s_{j}\quad\text{\text{if}\quad}i\equiv j\mod{K}.
\end{eqnarray}
Now suppose that the linear system has a valid solution $\{\lambda_{i}\in\mathcal{N}_0 : i\in\mathcal{Z}_{K(K+1)}\}$.
Without loss of generality, we assume $s_{0}=\min\{s_{k}:k\in\mathcal{Z}_{K}\}$.
To be more illustrative, we can rewrite \eqref{eq:K.user.lambda}
into the following matrix form
\begin{equation}
\left[\begin{array}{cccccccccc}
\lambda_{K^{2}} & \lambda_{K^{2}+1} & \cdots & \lambda_{0}\\
 & \lambda_{K^{2}+1} & \cdots & \lambda_{0} & \lambda_{1}\\
 &  &  &  & \vdots\\
 &  &  &  & \lambda_{1} & \lambda_{2} & \cdots & \lambda_{K+1}\\
 &  &  &  &  & \lambda_{2} & \cdots & \lambda_{K+1} & \lambda_{K+2}\\
 &  &  &  &  &  & \cdots &  &  & \cdots
\end{array}\right]\left[\begin{array}{c}
1\\
1\\
1\\
\vdots
\end{array}\right]=\left[\begin{array}{c}
s_{0}\\
s_{1}\\
\vdots\\
s_{K+1}\\
s_{K+2}\\
\vdots
\end{array}\right].
\end{equation}
From this illustrative form, we get $\lambda_{K^{2}+1}+\cdots+\lambda_{K(K+1)-1}+\lambda_{0}=s_{0}-\lambda_{K^{2}}\leq s_{0}$.
By substituting this inequality into $\lambda_{1}=s_{1}-(\lambda_{K^{2}+1}+\cdots+\lambda_{K(K+1)-1}+\lambda_{0})$,
we get
\begin{equation}
\lambda_{1}\geq s_{1}-s_{0}.
\end{equation}
Then applying it into $\lambda_{2}+\cdots+\lambda_{K+1}=s_{K+1}-\lambda_{1}$,
along with $s_{K+1}=s_{1}$, gives
\begin{equation}
\lambda_{2}+\cdots+\lambda_{K+1}\leq s_{0}.
\end{equation}
Sequentially applying it into $\lambda_{K+2}=s_{K+2}-(\lambda_{2}+\cdots+\lambda_{K+1})$,
together with $s_{K+2}=s_{2}$, we get
\begin{equation}
\lambda_{K+2}\geq s_{2}-s_{0}.
\end{equation}
Continuing this process, we can get
\begin{equation}
\lambda_{(i-1)K+i}\geq s_{i}-s_{0},\quad1\leq i\leq K-1.
\end{equation}
Due to the cyclic property of the linear system, we can easily see
that $\lambda_{i}$ should have the same property as $\lambda_{(i-1)K+i}$.
Therefore,
\begin{equation}
\lambda_{i}\geq s_{i}-s_{0},\quad1\leq i\leq K-1.
\end{equation}
Finally applying them into $\lambda_{0}+\lambda_{K}=s_{K}-(\lambda_{1}+\cdots+\lambda_{K-1})$,
together with $\lambda_{0}+\lambda_{K}\geq0$ and $s_{K}=s_{0}$,
we get $s_{0}\geq\sum_{i=1}^{K-1}(s_{i}-s_{0})$, or equivalently,
\begin{equation}
\sum_{k=0}^{K-1}s_{k}\leq(K+1)s_{0},
\end{equation}
which proves the condition given by \eqref{eq:K.user.SN.cond} is
a necessary condition.

\textbf{Sufficient condition}: We prove the condition is a sufficient
one by showing a valid solution for the linear system. As previously,
we assume $s_{0}=\min\{s_{k}:k\in\mathcal{Z}_{K}\}$. The valid solution
is given by
\begin{equation}
\lambda_{i}=\begin{cases}
s_{i}-s_{0}, & \text{if}\quad i\neq(j-1)K+j,j\in\{1,2,\cdots,K-1,K\}\\
(K-1)s_{0}-\sum_{k\in\{1,2,\cdots,K-1\},k\neq j}s_{k}, & \text{if}\quad i=(j-1)K+j,j\in\{1,2,\cdots,K-1\}\\
Ks_{0}-\sum_{k=0}^{K-1}s_{k} & \text{if}\quad i=K^{2}
\end{cases}
\end{equation}
where $s_{i}=s_{j}$ if $i\equiv j\mod{K}$, $j\in\mathcal{Z}_{K}$.
It is easy to prove in this solution set, $\lambda_{i}\ge0$ for all
$i\in\mathcal{Z}_{K(K+1)}$.
\end{IEEEproof}
\begin{rem}
As an example, we show a valid solution for $K=4$ as shown in \eqref{eq:solution.K4}. Also note that the solution is not unique.
\end{rem}

\begin{figure*}

\begin{equation}
\left[\begin{array}{c}
\lambda_{0}\\
\lambda_{1}\\
\lambda_{2}\\
\lambda_{3}\\
\lambda_{4}\\
\lambda_{5}\\
\lambda_{6}\\
\lambda_{7}\\
\lambda_{8}\\
\lambda_{9}\\
\lambda_{10}\\
\lambda_{11}\\
\lambda_{12}\\
\lambda_{13}\\
\lambda_{14}\\
\lambda_{15}\\
\lambda_{16}\\
\lambda_{17}\\
\lambda_{18}\\
\lambda_{19}
\end{array}\right]=\left[\begin{array}{c}
0\\
3s_{0}-(s_{2}+s_{3})\\
s_{2}-s_{0}\\
s_{3}-s_{0}\\
0\\
s_{1}-s_{0}\\
3s_{0}-(s_{1}+s_{3})\\
s_{3}-s_{0}\\
0\\
s_{1}-s_{0}\\
s_{2}-s_{0}\\
3s_{0}-(s_{1}+s_{2})\\
0\\
s_{1}-s_{0}\\
s_{2}-s_{0}\\
s_{3}-s_{0}\\
4s_{0}-(s_{1}+s_{2}+s_{3})\\
s_{1}-s_{0}\\
s_{2}-s_{0}\\
s_{3}-s_{0}
\end{array}\right]\label{eq:solution.K4}
\end{equation}

\end{figure*}

\section{Results and discussions}

\begin{figure}
\begin{centering}
\includegraphics{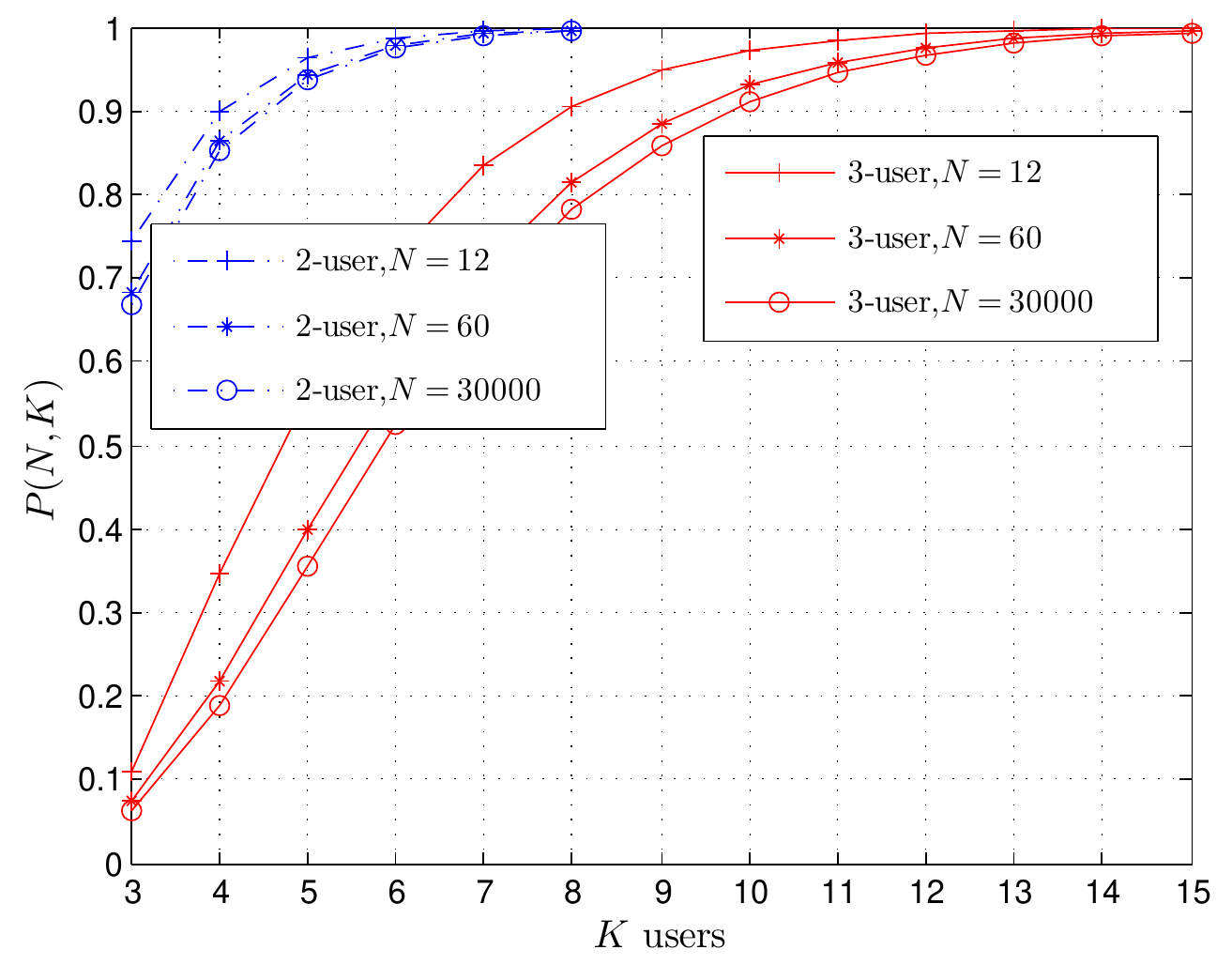}
\par\end{centering}
\caption{The probability $P(N,K)$ versus $K$ users for BIA-feasible $2$-user
BC and BIA-feasible $3$-user BC.}
\label{fig:PNK}
\end{figure}

Now, we show how the knowledge of BIA-feasibility condition can help improve the achievable DoF for a general homogenous $2\times 1$ BC. We show in Fig.~\ref{fig:PNK} the successful probability of finding, from $K$ homogeneous users, a BIA-feasible 2-user BC or a BIA-feasible 3-user BC. The probability rate of finding a BIA-feasible 3-user BC is derived from \eqref{eq:PNK3}, whileas the probability rate for a BIA-feasible 2-user BC is calculated by using
\begin{equation}
f(N,K,2) = 1+\sum_{n=2}^{N/3}\left( 2[n^{K-1}-(n-1)^{K-1}] +
(n-2)\gamma(n,K-1,2)\right).
\end{equation}
Note that the counterpart of $f(N,K,2)$ given in \cite{Zhou2012} is a lower bound, in which how the balls are placed at the two ends of the arc containing $n_{\delta,1}$ is not fully examined.

As shown in this figure, the successful probability drops as the coherent time $N$ increases, but the decreasing becomes negligible when $N$ is big, say $N\geq 60$ for both 2-user and 3-user settings. This observation indicates that the successful rate for finding BIA-feasible $k\times 1$ BC, $k=2,3$, converges very fast over $N$, and the $P(30000,K)$ can be regarded as the asymptotic/limit probability $P(\infty,K)$. The figure also shows that the successful rate increases with the user size $K$, and it is larger than $95\%$ when $K\geq 5$ and $K\geq 11$ for finding a BIA-feasible 2-user BC and a BIA-feasible 3-user BC, respectively. This implies that a $2\times 1$ BC network with homogeneous $K$ users can achieve $4/3$ DoF almost surely when $5\leq K <11$ by finding a BIA-feasible 2-user BC, and can achieve $3/2$ DoF almost surely when $11\leq K$ by finding a BIA-feasible 3-user BC.

From this figure, we can also roughly estimate, for different $K$, the expected DoF achieved by using BIA. For instance, on the range of $2\leq K\leq 4$, the successful probability of finding a BIA-feasible $2$-user BC, which provides 4/3 DoF, is larger than $50\%$ on average, so the expected DoF is $\tfrac{1}{2}(1+\tfrac{4}{3}) = \tfrac{7}{6}$ for this range of $K$. Similarly, for the range of $5\leq K<11$, the expected DoF is given by $\tfrac{1}{2}(\tfrac{4}{3}+\tfrac{3}{2})=\tfrac{17}{12}$. As the achievable DoF is $\tfrac{2K}{2+K-1}$ for a BIA-feasible $K$-user $2\times 1$ BC, it is evident that the asymptotic expected DoF is 2, which is the maximal DoF available by a $2\times K$ MIMO channel. Therefore a homogenous $2\times 1$ BC with $K$ users asymptotically forms a virtual $2\times K$ MIMO channel from the DoF perspective.

\section{Conclusion}
In this paper we examined the BIA-feasibility problem in a $K$-user $2\times 1$ BC with homogeneous block fading. By casting the problem into the solvability problem of a system of linear Diophantine equations, we find the sufficient and necessary condition on the block offsets $(n_{\delta,1},\cdots,n_{\delta,K})$ for BIA to achieve the optimal $\tfrac{2K}{2+K-1}$ DoF. We also provide solutions to achieve the optimal DoF. The analysis method proposed in this paper offers a potential tool to study the BIA-feasibility problem for a MISO BC with more general heterogeneous block fading, which is one of our ongoing research topics.

Based on the BIA-feasible condition derived above and the justified assumption that all users' fading blocks are independently and uniformly placed, we further studied the probability of finding a BIA-feasible $3$-user MISO BC when there are $K\geq 3$ homogeneous users. The numerical analysis shows that it is almost sure (more than $95\%$ certainty) to find such a BIA-feasible $3$-user MISO BC if $K\geq 11$. It is also evident that a homogenous $K$-user $2\times 1$ BC achieves the optimal 2 DoF by using BIA when $K$ goes large, forming a virtual $2\times K$ MIMO channel.


\end{document}